\documentclass[british,a4paper,10pt]{article}

\usepackage{amssymb}
\usepackage{amsmath}
\usepackage{subfigure}
\usepackage{slashed}
\usepackage{amsmath}
\usepackage[normalem]{ulem}
\usepackage{graphicx,psfrag}
\usepackage{mathrsfs}
\usepackage{amsfonts}
\usepackage{multirow}
\usepackage{hyperref}
\usepackage{placeins}
\usepackage{amsthm}
\usepackage{latexsym}
\usepackage[dvips]{epsfig}
\usepackage{enumitem}  
\usepackage{soul}
\usepackage{graphicx}
\usepackage{amsthm}
\usepackage{latexsym}
\usepackage[dvips]{epsfig}
\usepackage[utf8]{inputenc}
\usepackage[english]{babel}
\usepackage{multicol}
\usepackage{color}
\usepackage{comment}
\usepackage[dvipsnames]{xcolor}
\usepackage{wasysym}
\usepackage{mathrsfs}
\usepackage{eufrak}
\usepackage{bm}
\usepackage{authblk}
\usepackage{slashed}
\usepackage{yhmath} 
\usepackage{stmaryrd}

\theoremstyle{plain}
\newtheorem{proposition}{Proposition}
\newtheorem{lemma}{Lemma}

\newtheorem{remark}{Remark}

\def\bmeta{{\bm \eta}}
\def\bmg{{\bm g}}
\def\bmsigma{{\bm \sigma}}
\def\uell{\underline{\ell}}
\def\uL{\underline{L}}
\def\bme{{\bm e}}
\def\bmp{{\bm p}}

\allowdisplaybreaks
\begin{document}

%%%%%%%%%%%%%%%%%%%%%%%%%%%%%%%%%%%%%%%%%%%%%%%%%%%%%%%%%%%%%
\title{\textbf{The good-bad-ugly system near spatial infinity on flat
    spacetime}}
%%%%%%%%%%%%%%%%%%%%%%%%%%%%%%%%%%%%%%%%%%%%%%%%%%%%%%%%%%%%%

\author[1,2]{Miguel Duarte\footnote{E-mail address:{\tt
   miguelcsduarte@tecnico.ulisboa.pt }}}
\author[,2]{Justin Feng
  \footnote{E-mail address:{\tt justin.feng@tecnico.ulisboa.pt}}}
\author[,2]{Edgar Gasper\'in
  \footnote{E-mail address:{\tt edgar.gasperin@tecnico.ulisboa.pt}}}
\author[,2]{David Hilditch
   \footnote{E-mail address:{\tt david.hilditch@tecnico.ulisboa.pt}}}
%%%%%%%%%%%%%%%%%%%%%%%%%%%%%%%
\affil[1] {CAMGSD, Departamento de Matem\'atica, Instituto Superior T\'ecnico IST,
  Universidade de Lisboa UL, Avenida Rovisco Pais 1, 1049 Lisboa, Portugal}
  \affil[2] {CENTRA, Departamento de F\'isica, Instituto Superior
  T\'ecnico IST, Universidade de Lisboa UL, Avenida Rovisco Pais 1,  1049 Lisboa, Portugal}
%%%%%%%%%%%%%%%%%%%%%%%%%%%%%%%

\maketitle
  
\begin{abstract}
A model system of equations that serves as a model for the Einstein
field equation in generalised harmonic gauge called the good-bad-ugly
system is studied in the region close to null and spatial infinity in
Minkowski spacetime. This analysis is performed using H.~Friedrich's
cylinder construction at spatial infinity and defining suitable
conformally rescaled fields. The results are translated to the
physical set up to investigate the relation between the
polyhomogeneous expansions arising from the analysis of linear fields
using the~$i^0$-cylinder framework and those obtained through a
heuristic method based on H\"ormander's asymptotic system.
\end{abstract}


%%%%%%%%%%%%%%%%%%%%%%%%%%%%%%%%%%%%%%%%%%%%%%%%%%%%%%%%%%%%%
\section{Introduction}
%%%%%%%%%%%%%%%%%%%%%%%%%%%%%%%%%%%%%%%%%%%%%%%%%%%%%%%%%%%%%

One of the most emblematic results in the classical theory of
asymptotics in general relativity is the peeling theorem~\cite{Sac61,
  BonBurMet62, NewPen62}. The general term of ``peeling'' refers to
the decay of the Weyl tensor in the asymptotic region of the
spacetime. The classical peeling theorem~\cite{NewPen62} shows that,
if a spacetime admits a \emph{smooth conformal extension} then the
components of the Weyl tensor decay as integer powers of a suitable
parameter along the generators of outgoing light cones. The genericity
of this crucial smoothness assumption has been put in question from
the perspective of the initial value problem. There exist a
considerable number of results ---\cite{DuaFenGas22, Lin17,
  ChrMacSin95, Win85, Keh21, GasVal18, Val99, Fri98a}--- showing, with
varying levels of rigour and different standpoints, that generically,
the gravitational field would satisfy at best a restricted peeling
behaviour. Comparing these different results becomes a non-trivial
task due to the diverse approaches taken in each case and the variety
of gauges used to derive the results.

Recently, in~\cite{DuaFenGas22} it was shown, exploiting a heuristic
method introduced in~\cite{DuaFenGasHil21}, that in generalised
harmonic gauge, the components of the Weyl tensor admit a
polyhomogeneous expansion.  The heuristic method put forward
in~\cite{DuaFenGasHil21}, is based on a generalisation of
H\"ormander's asymptotic system ---see~\cite{LinRod03, GasHil18,
  Kei17}.  The general method used in~\cite{DuaFenGasHil21} finds
applicability and is tailored for a formulation of the Einstein field
equations in generalised harmonic gauge which is designed for
numerical investigations via suitably hyperboloidal initial value
problems exploiting the dual frame formalism~\cite{DuaFenGasHil22a,
  Hil15}.

Interestingly, there exist other body of work ---see \cite{Fri98a,
  Fri03, Val07, GasVal18, Fri18}--- based on a distinctively different
approach, the conformal Einstein field equations, giving seemingly
similar polyhomogeneous expansions for the Weyl tensor.  The
polyhomogeneous expansions in~\cite{Fri98a, Val07, GasVal18} are
obtained through the analysis of the components of the \emph{rescaled
Weyl spinor} close to spatial and null infinity. To do so, the
framework of the \emph{cylinder at spatial infinity} is employed. The
cylinder at spatial infinity is a formalism developed to study the
behaviour of the gravitational field in the region where null and
spatial infinity meet ---the critical sets--- by means of the
\emph{extended conformal Einstein field equations} written in a gauge
adapted to a special class of curves known as conformal
geodesics~\cite{Fri98a, Val16}.  This special gauge around which the
$i^0$-cylinder formalism is constructed is known as the $F$-gauge.
Unfortunately, the relation between the $F$-gauge and other more
traditional gauges is not simple to establish in general.  However,
there is a particular case where this gauge and the construction of
the cylinder at spatial infinity can be obtained in an explicit closed
form: the Minkowski spacetime.  This special conformal representation
of the Minkowski spacetime has been used as a model to understand the
behaviour of fields propagating in the vicinity of spatial infinity of
asymptotically flat spacetimes and the consequences of the degeneracy
of the equations in the critical sets ---see \cite{Fri02, Val03a,
  GasVal20, GasVal21a, ValAli22, BeyDouFra13,
  BeyDouFra13a}. Crucially, one of these consequences is that even
linear fields propagating in this conformal representation of the
Minkowski spacetime ($i^0$-cylinder background) will develop
logarithmic terms at the critical sets which spread out to null
infinity ---see~\cite{Fri02, Val03a, GasVal21a, MinMacVal22}.
Although it has been shown that the non-linearities in the conformal
Einstein field equations will generate further logarithmic terms
---see~\cite{Val04}, the analysis of linear fields about the given
background already serves as a basis to develop intuition into what is
the minimal regularity of the field that can be be expected close to
the critical sets.

On the other hand, the basic model under which the general method
of~\cite{DuaFenGasHil21} was constructed is the good-bad-ugly (GBU)
system of equations.  This system is constituted by three fields which
satisfy wave equations with non-linearities that mimic the worst of
those present in the Einstein field equations in generalised harmonic
gauge. In this paper, the GBU model equation on flat spacetime is
solved using the methods of the cylinder at spatial infinity. The
solution is obtained by conformally transforming the equations and
fields to the $i^0$-cylinder background, defining a corresponding
``unphysical'' GBU system, then solving for the unphysical good, bad
and ugly fields close to spatial infinity and then translating back
the solution to the physical picture and comparing the results with
those obtained in~\cite{DuaFenGasHil21}.  In doing so, we clarify the
relation between the logarithmic terms obtained through the two
methods and establish a base analysis for future investigations in the
non-linear case.

%%%%%%%%%%%%%%%%%%%%%%%%%%%%%%%%%%%%%%%%%%%%%%%%%%%%%%%%%%%%%
\section*{Notations and Conventions}
\label{NotationConventions}
%%%%%%%%%%%%%%%%%%%%%%%%%%%%%%%%%%%%%%%%%%%%%%%%%%%%%%%%%%%%%

Most of the literature of Friedrich's cylinder at spatial infinity
uses the signature convention natural to
spinors~$(+,-,-,-)$. Nonetheless, since spinor formalism will not be
used in this article, the signature convention for a Lorentzian
spacetime metric will be the more standard $(-,+,+,+)$.  Fields
defined on the physical setup can be identified by the $\;\tilde{}\;$
symbol, while the unphysical (compactified) ones will not have such
decoration. Latin and Greek indices will be used as abstract and
coordinate indices respectively.

%%%%%%%%%%%%%%%%%%%%%%%%%%%%%%%%%%%%%%%%%%%%%%%%%%%%%%%%%%%%%
\section{The cylinder at spatial infinity}\label{TheCylinderAtSpatialInfinity}
%%%%%%%%%%%%%%%%%%%%%%%%%%%%%%%%%%%%%%%%%%%%%%%%%%%%%%%%%%%%%
 
The term cylinder at spatial infinity is broadly used to refer to a
general framework to study conformal extensions of asymptotically flat
spacetimes in a neighbourhood of spatial infinity using the conformal
Einstein field equations ---see~\cite{Fri98a, Val16}. While these
conformal extensions are known in general only in an abstract
(non-explicit) way, for the Minkowski spacetime one can write closed
and explicit expressions. Although most of the expressions given in
this section have been reported already ---see for
instance~\cite{Fri98a, FriKan00, Fri02, GasVal20, GasVal21a,
  MinMacVal22}--- in the discussion presented here the emphasis is
placed on making contact with the \emph{physical} structures and the
translation of some of the~$i^0$-cylinder results to the physical
set-up.

%%%%%%%%%%%%%%%%%%%%%%%%%%%%%%%%%%%%%%%%%%%%%%%%%%%%%%%%%%%%%
\subsection{The $i^0$-cylinder conformal extension of the
  Minkowski spacetime}\label{i0MinkowskiGeneral} 
%%%%%%%%%%%%%%%%%%%%%%%%%%%%%%%%%%%%%%%%%%%%%%%%%%%%%%%%%%%%%

Let $\tilde{\bmeta}$ be a Minkowski metric and let $\tilde{x}^{\mu }=
(\tilde{t},\tilde{x}^{i})$, denote \emph{physical} Cartesian
coordinates. In these coordinates the Minkowski metric reads:
\begin{align}
\tilde{\bmeta}=
\eta_{\mu\nu}\mathbf{d}\tilde{x}^{\mu}\otimes\mathbf{d}\tilde{x}^{\nu},
\end{align}
where~$\eta_{\mu\nu}=\text{diag}(-1,1,1,1)$. Introducing
\emph{physical} polar coordinates defined by~$\tilde{\rho}^2=
\delta_{ij}\tilde{x}^{i}\tilde{x}^{j}$
where~$\delta_{ij}=\text{diag(1,1,1)}$ and an arbitrary choice of
coordinates on~$\mathbb{S}^2$ one has
\begin{align}\label{MinkowskiMetricPhysicalPolar}
\tilde{\bmeta}=-\mathbf{d}\tilde{t}\otimes\mathbf{d}\tilde{t}
+\mathbf{d}\tilde{\rho}\otimes \mathbf{d}\tilde{\rho}+\tilde{\rho}^2
\mathbf{\bm\sigma},
\end{align}
with~$\tilde{t}\in(-\infty, \infty)$, $\tilde{\rho}\in [0,\infty)$
  where~$\bm\sigma$ denotes the standard metric on~$\mathbb{S}^2$.
  The procedure to obtain the cylinder representation of spatial
  infinity is as follows: First, introduce \emph{inversion Cartesian
  coordinates}~$x^{\mu}=(t,x^{i})$
 \begin{align}\label{UnphysicalCartesianToPhysicalCartesian}
x^{\mu}={\tilde{x}^{\mu}}/{\tilde{X}^2}, \qquad \tilde{X}^2 \equiv
\tilde{\eta}_{\mu\nu}\tilde{x}^{\mu}\tilde{x}^{\nu}.
 \end{align}
A direct calculation shows that the inverse transformation is
\begin{align}\label{PhysicalCartesianToUnphysicalCartersian}
\tilde{x}^{\mu}=x^{\mu}/X^2, \qquad X^2=\eta_{\mu\nu}x^{\mu}x^{\nu}.
\end{align}
which is valid in the region where $\tilde{X}^2>0$, namely, in the
complement of the lightcone at the origin in the Minkowski spacetime.
 \begin{remark}
   \emph{There is a sign difference in equation
   \eqref{UnphysicalCartesianToPhysicalCartesian} respect to other
   discussions in the literature ---see \cite{GasVal20, GasVal21a,
     MinMacVal22}--- which is due to the use of a different signature
   convention.}
 \end{remark}
Using these coordinates, the following conformal (inversion) extension
of the Minkowski spacetime is identified
\begin{align}
\label{InverseMinkowskiMetricDef}
\bmg_{I}=\Xi^2 \hspace{0.5mm}\tilde{\bmeta},
\end{align}
where~$\bmg_{I}=\eta_{\mu\nu}\mathbf{d}x^{\mu}\otimes
\mathbf{d}x^{\nu}$ and $\Xi =X^2$.  Observe that~$X^2=1/\tilde{X}^2$.
Now, define an \emph{unphysical polar radial coordinate} via
$\rho^2=\delta_{ij}x^{i}x^{j}$. In the unphysical polar coordinates,
the rescaled metric~$\bmg_{I}$ and conformal factor~$\Xi$ read
\begin{align}
\label{InverseMinkowskiUnphysicaltrhocoords}
\bmg_{I}=-\mathbf{d}t\otimes\mathbf{d}t +\mathbf{d}\rho\otimes
\mathbf{d}\rho+\rho^2 \mathbf{\bm\sigma}, \qquad \Xi=\rho^2 -t^2,
\end{align}
with~$t\in(-\infty,\infty)$
and~$\rho\in~[0,\infty)$. Although~$\bmg_{I}$ is again the Minkowski
  metric notice that the roles of the origin~$\mathcal{O}$ and spatial
  infinity~$i^0$ are swapped. In other words, in this
  representation,~$i^0$ corresponds to the point with
  coordinates~$(t=0,\rho=0)$ in $(\mathbb{R}^4,\bmg_{I})$. To prepare
  for the upcoming discussion, it will be useful to write the
  coordinate
  transformations~\eqref{UnphysicalCartesianToPhysicalCartesian}
  and~\eqref{PhysicalCartesianToUnphysicalCartersian} in terms of the
  physical and unphysical (inversion) polar coordinates:
\begin{align}
 \label{UnphysicalToPhysicaltrho}
  \tilde{t}=\frac{t}{\rho^2-t^2}, \qquad \tilde{\rho}=
  \frac{\rho}{\rho^2-t^2}.
\end{align}
The inverse transformation is given by
\begin{align}
 \label{physicalToUnphysicaltrho}
   t=\frac{\tilde{t}}{\tilde{\rho}^2-\tilde{t}^2}, \qquad \rho =
   \frac{\tilde{\rho}}{\tilde{\rho}^2-\tilde{t}^2}.
\end{align}
where we have chosen the positive root in the expression defining the
respective radial coordinates.  To arrive to the relevant conformal
representation of the Minkowski spacetime the following change of
coordinates is introduced
\begin{align}
  t=\rho\tau.
\end{align}
The unphysical coordinate system $(\tau,\rho)$ will be
called~$F$-coordinates.  In these coordinates the metric~$\bmg_{I}$ is
written as
\begin{align*}
\bmg_{I}=-\rho^2 \mathbf{d}\tau\otimes \mathbf{d}\tau
+(1-\tau^2)\mathbf{d}\rho \otimes \mathbf{d}\rho - \rho\tau
\mathbf{d}\rho\otimes \mathbf{d}\tau - \rho\tau \mathbf{d}\tau \otimes
\mathbf{d}\rho + \rho^2 \bmsigma.
\end{align*}
Finally, by rescaling the $\bmg_{I}$ as
\begin{align}
 \label{MetricCylinderToIversedMinkowski}
\bmg \equiv \frac{1}{\rho^2} \bmg_{I},
\end{align}
one obtains the conformal representation of the Minkowski spacetime
adapted to the cylinder at spatial infinity.  Composing the coordinate
transformations and the conformal transformations one concludes that
the relation between the (physical) Minkowski metric~$\tilde{\bm\eta}$
and the (unphysical)~$i^0$-cylinder metric~$\bmg$ is given by
\begin{align}
 \bmg=\Theta^2\tilde{\bm\eta}
\end{align}
where
\begin{align*}
\bmg=- \mathbf{d}\tau\otimes \mathbf{d}\tau
+\frac{(1-\tau^2)}{\rho^2}\mathbf{d}\rho \otimes \mathbf{d}\rho -
\frac{\tau}{\rho} \mathbf{d}\rho\otimes \mathbf{d}\tau -
\frac{\tau}{\rho} \mathbf{d}\tau \otimes \mathbf{d}\rho + \bmsigma.
\end{align*}
and
\begin{align}
\Theta := \frac{\Xi}{\rho}
\end{align}
consistent with the bookkeeping naming conventions of
Section~\ref{NotationConventions}, the metric~$\bmg$ will be
regarded as the metric of the \emph{unphysical spacetime}
while~$\tilde{\bm\eta}$ as the metric of the \emph{physical
spacetime}. These naming conventions stem from the fact that the
unphysical metric~$\bmg$ corresponds to an explicit solution to the
conformal Einstein field equations written in a gauge adapted to
conformal geodesics ---see \cite{Fri98a, Val16, FriKan00, Fri02,
  GasVal20, GasVal21a, MinMacVal22} for further discussion and
definitions. To simplify the terminology when discussing fields
propagating in this geometry we will refer to it simply as
the~$i^0$-cylinder background.
To better describe the geometry of the cylinder at spatial infinity it
is convenient to introduce, the following~$\bmg$-null frame ---which
in the following will be referred to as the~$F$-frame:
\begin{align}\label{Fframe}
 \bm\ell
  =(1+\tau)\bm\partial_{\tau}{} -
  \rho\bm\partial_{\rho}{},  \qquad
  \bm\uell =(1-\tau)\bm\partial_{\tau}{} +
  \rho\bm\partial_{\rho}{}, \qquad
  \bm\partial_{+}{} , \qquad
  \qquad
  \bm\partial_{-}{}. 
\end{align}
The corresponding dual coframe is given by
\begin{align*}
&  \bm\ell^{\flat}=-\mathbf{d}\tau -
  \frac{1}{\rho}\big(1+\tau\big)\mathbf{d}\rho,  \qquad
  \bm\uell^{\flat}=-\mathbf{d}\tau{}
  +\frac{1}{\rho}\big(1-\tau\big)\mathbf{d}\rho,\qquad
  \bm\omega^{+}, \qquad
  \bm\omega^{-}.
\end{align*}
In these expressions~$\bm\partial_\pm$ and~$\bm\omega^\pm$ represent
an arbitrary null frame and coframe on~$\mathcal{Q}\approx
\mathbb{S}^2$, denoting the surfaces of constant~$\tau$ and
constant~$\rho$ so that
\begin{align}
\bm\sigma=2(\bm\omega^{+}\otimes
 \bm\omega^{-}+\bm\omega^{-}\otimes \bm\omega^{+}), \qquad
 \bm\sigma^{\flat}= \frac{1}{2}(\bm\partial_{+}\otimes \bm\partial_{-}
 + \bm\partial_{-}\otimes \bm\partial_{+}).
\end{align}
This frame is Lie dragged along the~$\bm\partial_\tau$
and~$\bm\partial_\rho$ directions, imposing
that~$[\bm\partial_{\tau},\bm\partial_{\pm}]=
[\bm\partial_{\rho},\bm\partial_{\pm}]=0$ ---see discussion in
Appendix of~\cite{GasVal20}.  In accordance with the conventions of
this article, the metric then reads
\begin{align}\label{UnphysicalMetricNullTetrad}
g_{ab}=\ell_{(a}\uell_{b)} -  \omega^{+}_{(a}\omega^{-}_{b)}\,,
\end{align}
so that the normalisation of the tetrad
is~$\ell_a\uell^a=-\omega^{+}_a\partial_{-}^a=-2$ while all the other
contractions vanish. The relevance of the geometry of the cylinder at
spatial infinity for Minkowski spacetime, is that, in this
representation future and past null infinity are located at
\begin{align}
 \mathscr{I}^{+} \equiv \{ p \in \mathcal{M} \; \rvert\; \tau(p) =1
 \}, \qquad \mathscr{I}^{-} \equiv \{ p \in \mathcal{M} \; \rvert \;
 \tau(p) =-1\},
 \end{align}
and the following sets can be distinguished:
\begin{align*}
 I \equiv \{ p \in \mathcal{M} \; \rvert \;\; |\tau(p)|<1, \;
 \rho(p)=0\}, \qquad I^{0} \equiv \{ p \in \mathcal{M}\; \rvert \;
 \tau(p)=0, \; \rho(p)=0\},
\end{align*} 
\begin{align*}
 I^{+} \equiv \{ p\in \mathcal{M} \; \rvert \; \tau(p)=1, \; \rho(p)=0
 \}, \qquad I^{-} \equiv \{p \in \mathcal{M}\; \rvert \; \tau(p)=-1,
 \; \rho(p)=0\}.
\end{align*}
From the last expressions it can be noticed that spatial
infinity~$i^0$ has been blown up to the cylinder $I$.  Moreover,
different cuts of~$I$ can be identified:~$I^0$ denotes the
intersection of the time symmetric hypersurface~$\tau=0$ and~$I$ and
the critical sets~$I^{\pm}$ represent the region where spatial and
future/past null infinity meet, where evolution PDEs typically
degenerate ---see~\cite{Fri98a, Fri02, GasVal20, GasVal21a,
  MinMacVal22}.

%%%%%%%%%%%%%%%%%%%%%%%%%%%%%%%%%%%%%%%%%%%%%%%%%%%%%%%%%%%%%
\subsection{Relation to the physical coordinates and physical frame}
%%%%%%%%%%%%%%%%%%%%%%%%%%%%%%%%%%%%%%%%%%%%%%%%%%%%%%%%%%%%%

The relation between the $F$-coordinates and the physical polar
coordinates is:
\begin{align}\label{Ftophys}
  \tau = \frac{\tilde{t}}{\tilde{\rho}}, \qquad \rho =
  \frac{\tilde{\rho}}{\tilde{\rho}^2-\tilde{t}^2},
\end{align}
and the inverse transformation is
\begin{align}\label{phytoF}
  \tilde{t} =
  \frac{\tau}{\rho (1-\tau^2)}, \qquad
  \tilde{\rho}=\frac{1}{\rho (1-\tau^2)}.
\end{align}
Unwrapping the definitions, the conformal factor~$\Theta$
in~$F$-coordinates and physical coordinate respectively, reads
\begin{align}
  \Theta = \rho (1-\tau^2) = \frac{1}{\tilde{\rho}}
\end{align}
Hence, the relations in \eqref{phytoF} can be succinctly rewritten as
\begin{align}
  \tilde{t}=\frac{\tau}{\Theta}, \qquad \tilde{\rho}=
  \frac{1}{\Theta}.
\end{align}
For the upcoming calculations it will be convenient to introduce the
physical advanced and retarded times
\begin{align}
\tilde{u}=\tilde{t}- \tilde{\rho}, \qquad \tilde{v}= \tilde{t}+
\tilde{\rho}
\end{align}
The associated physical null vectors
\begin{align}\label{PhysicalFrame}
L^a =
-\tilde{\eta}^{ab}\tilde{\nabla}_{b}\tilde{u}, \qquad
\uL^a =
-\tilde{\eta}^{ab}\tilde{\nabla}_{b}\tilde{v},
\end{align}
explicitly read
\begin{align}\label{PhysicalFrameExplicit}
L = \bm\partial_{\tilde{t}}
+\bm\partial_{\tilde{\rho}}, \qquad \uL =
\bm\partial_{\tilde{t}}-\bm\partial_{\tilde{\rho}}.
\end{align}
These vectors can be complemented with a pair of complex null vectors
so that the physical null frame~$\{L, \; \uL,\; \tilde{\bm\omega}^{+},
\tilde{\bm\omega}^{-} \}$ reads
\begin{align}
L =  \bm\partial_{\tilde{t}} +\bm\partial_{\tilde{\rho}}, \qquad \uL =
\bm\partial_{\tilde{t}}-\bm\partial_{\tilde{\rho}},
\qquad
\tilde{\omega}^{+}=\tilde{\rho}^{-1}\bm\partial_{+},
\qquad
\tilde{\omega}^{-}=\tilde{\rho}^{-1}\bm\partial_{-},
\end{align}
with the normalisation understood to be taken with respect to the
physical Minkowski metric $\tilde{\bmeta}$
\begin{align}
\tilde{\eta}_{ab}L^a\uL^b
=-\tilde{\eta}_{ab}\tilde{\omega}^{a}_{+}\tilde{\omega}^{a}_{-}
=-2.
\end{align}
This normalisation implies that the physical metric can be written as
\begin{align}\label{PhysicalMetricNullTetrad}
 \tilde{\eta}_{ab}=L_{(a}\uL_{b)} -  \tilde{\omega}^{+}_{(a}\tilde{\omega}^{-}_{b)}.
\end{align}
\begin{remark}
  \emph{ Consistent with the previously stated conventions to
  distinguish the physical vs the unphysical fields, the physical null
  vectors should be denoted by the symbols $\tilde{\bm\ell}$ and
  $\tilde{\underline{\bm\ell}}$. Nonetheless, to simplify the notation
  and to align it with conventions of \cite{GasVal18,DuaFenGasHil21,
    LinRod03}, the physical outgoing and incoming null vectors have
  been denoted with $L$ and $\underline{L}$ instead.  Observe that
  with the current conventions an orthonormal frame can be constructed
  so that the timelike legs of such tetrads would
  read~$\tilde{\bme}_{0}=\frac{1}{2}(L + \uL
  )=\bm\partial_{\tilde{t}}$ and~${\bme}_{0}=\frac{1}{2}({\bm\ell} +
  {\bm\uell})=\bm\partial_{\tau}$.}
\end{remark}

For the calculations to follow, it will be necessary to spell out not
only the relation between the coordinates but also the relation
between the~$F$-frame and the physical frames. A straightforward
calculation using equations~\eqref{Fframe}, \eqref{PhysicalFrame},
\eqref{Ftophys} and~\eqref{phytoF} gives the following

\begin{proposition}
  \label{Prop:FframeToPhsyicalframe}
  The relation between the $F$-frame and the physical frame ---as
  given in equations \eqref{Fframe} and \eqref{PhysicalFrame}, in the
  $i^0$-cylinder conformal extension of the Minkowski spacetime is
  given by
\begin{align}
L = \Theta \kappa^{-1} \bm\ell, \qquad
\uL = \Theta
\kappa \bm\uell, \qquad \tilde{\bm\partial}_{\pm}^{a}=
\Theta \bm\partial_{\pm}^{a},
\end{align}
where the conformal factor in~$F$-coordinates and physical coordinates
reads
\begin{align}\label{CF-theta}
  \Theta := \rho (1-\tau^2) = \frac{1}{\tilde{\rho}}
\end{align}
and, similarly, the boost parameter~$\kappa$ is given by
\begin{align}\label{boostphysical}
\kappa := \frac{1+\tau}{1-\tau} = -\frac{\tilde{v}}{\tilde{u}}.
\end{align}
\end{proposition}

\begin{remark}
  \emph{The last result simply emphasises that the relation between
  the physical and unphysical frame is not only a conformal
  transformation but also a boost encoded in~$\kappa$. Observe
  that~$\kappa$ and~$\kappa^{-1}$ diverge at~$\mathscr{I}^-$
  and~$\mathscr{I}^+$ respectively; in particular, they diverge at the
  critical sets~$I^{\pm}$ but are well defined elsewhere in the
  cylinder~$I$. A crucial observation for the subsequent discussion is
  that the boost factor~$\kappa$ ---as given in equation
  \eqref{boostphysical}--- can be written as the quotient between the
  physical advanced and retarded times.  }
\end{remark}

Although the previous calculation is very simple, identifying the
Lorentz transformation between the frames is crucial for practical
applications of the~$i^0$-cylinder framework.  An example of this is
that clarifying the relation between the~$F$-frame and the NP-frame is
central to compute conserved quantities at~$\mathscr{I}$ ---see for
instance~\cite{MohVal22, FriKan00, GasVal20}.

%%%%%%%%%%%%%%%%%%%%%%%%%%%%%%%%%%%%%%%%%%%%%%%%%%%%%%%%%%%%%
\section{The GBU model close to spatial infinity}\label{ModelEquations}
%%%%%%%%%%%%%%%%%%%%%%%%%%%%%%%%%%%%%%%%%%%%%%%%%%%%%%%%%%%%%

In~\cite{DuaFenGasHil21}, through an approach based on H\"ormander's
asymptotic system, formal polyhomogeneous expansions near null
infinity were obtained for a class of model equations called
good-bad-ugly. The motivation for these model equations is that they
mimic the non-linearities found in the Einstein field equations in
harmonic gauge. Moreover, in~\cite{DuaFenGas22} these expansions were
used to obtain formal asymptotic expressions for the Weyl
scalars. These were then used to assess the peeling properties of the
gravitational field arising from an initial value problem using the
Einstein field equations in generalised harmonic gauge. On the other
hand, in~\cite{Fri98a, FriKan00, Fri18, GasVal18, Val04d} similar
looking expansions have been obtained for the rescaled Weyl spinor
using the conformal Einstein field equations. In this section we
analyse the model equations of~\cite{DuaFenGasHil21} from the point of
view of conformal methods. Specifically, exploiting the framework of
Friedrich's cylinder at spatial infinity to understand if the
logarithmic terms in~\cite{GasVal18, Val04d} and~\cite{DuaFenGasHil21}
---sourcing the violation of peeling in~\cite{DuaFenGas22}--- are
related or not. To do so, we perform this calculation not on the full
non-linear case of the Einstein field equations but on a simple
good-bad-ugly model.

The good-bad-ugly system consists of the following equations on the
physical Minkowski spacetime~$(\tilde{\mathcal{M}},\tilde{\bmeta})$:
\begin{subequations}\label{gbu}
\begin{align}
  & \tilde{\square} \tilde{\phi}_{g} = 0, \label{gbu-good}
  \\ &\tilde{\square} \tilde{\phi}_b = (\nabla_{\tilde{t}}
  \tilde{\phi}_g)^2 , \label{gbu-bad} \\ &\tilde{\square} \tilde{\phi}_u =
  \frac{2}{\tilde{\rho}}\nabla_{\tilde{t}} \tilde{\phi}_u. \label{gbu-ugly}
\end{align}
 \end{subequations}
Here~$\tilde{\square}
:=\tilde{\eta}^{ab}\tilde{\nabla}_a\tilde{\nabla}_b$
where~$\tilde{\nabla}$ is the Levi-Civita connection of
$\tilde{\bmeta}$. Here~$\tilde{\phi}_g$, $\tilde{\phi}_b$
and~$\tilde{\phi}_u$ are scalar fields that are called ~{\it good,
  bad} and {\it ugly} fields, respectively.  Our aim is to analyse
\eqref{gbu} using conformal methods and the framework of the cylinder
at spatial infinity. To do so, recall that for two conformally related
manifolds ---not necessarily the~$i^0$-cylinder and the Minkowski
spacetime--- $(\tilde{\mathcal{M}},\tilde{\bmg})$ and
$(\mathcal{M},\bmg)$ with $\bmg=\Omega^2\tilde{\bmg}$, the
D'Alembertian operator transforms under conformal transformations as
follows:
\begin{align}
  \square \phi- \frac{1}{6} \phi R = \Omega ^{-3}
  \bigg( \tilde{\square }\tilde{\phi}- \frac{1}{6}
\tilde{\phi} \tilde{R}\bigg), \label{General_Wave_Conformal_Transformation}
\end{align}
where~$R$ and~$\tilde{R}$ are the Ricci scalars of~$\bmg$
and~$\tilde{\bmg}$ respectively and $\phi = \Omega^{-1}\tilde{\phi}$.

%%%%%%%%%%%%%%%%%%%%%%%%%%%%%%%%%%%%%%%%%%%%%%%%%%%%%%%%%%%%%
\subsection{The good equation}
\label{Subsection:GoodEquation}
%%%%%%%%%%%%%%%%%%%%%%%%%%%%%%%%%%%%%%%%%%%%%%%%%%%%%%%%%%%%%

Using the conformal transformation formula for the wave equation given
in equation~\eqref{General_Wave_Conformal_Transformation},
substituting the wave equation~$\tilde{\square} \tilde{\phi}_{g} = 0$
on the physical Minkowski
spacetime~$(\tilde{\mathcal{M}},\tilde{\bmeta})$ and choosing the
target conformal extension ---the unphysical spacetime---
$(\mathcal{M},\bmg)$ to be that of Friedrich's cylinder at spatial
infinity discussed in section \ref{TheCylinderAtSpatialInfinity}, one
obtains
\begin{align}\label{unphysical_good}
  \square \phi_g =0.
\end{align}
To obtain the last equation we have used that in this very special
case~$R=\tilde{R}=0$. Notice that this means that the only non-zero
part of the Riemann curvature $R^a{}_{bcd}$ of the unphysical
spacetime is contained in the tracefree part of the (unphysical) Ricci
tensor $R_{\{ab\}}$.  Thus, the \emph{unphysical good} equation, is
simply the wave equation for the unphysical field
$\phi_g=\Theta^{-1}\tilde{\phi}_g$ propagating on the $i^0$-cylinder
background $(\mathcal{M},\bmg)$.  We stress that this true only for
this particular case since for a general conformal transformation $R$
does not necessarily vanish and the unphysical equation can become
potentially singular.

A direct calculation using the expressions given in
Section~\ref{TheCylinderAtSpatialInfinity} shows that the unphysical
good equation in the $F$-coordinates explicitly reads
\begin{align}\label{UnphysicalGoodEqExplicit}
  (\tau ^2-1) \partial _{\tau}^2 \phi -2 \rho \tau \partial
  _{\tau}\partial _{\rho}\phi +  \rho ^2 \partial _{\rho}^2\phi
  + 2 \tau \partial _{\tau}\phi +
  \Delta _{\mathbb{S}^{2}{}}{}\phi   = 0,
\end{align}
where~$\Delta _{\mathbb{S}^{2}{}}{}$ is the Laplace operator on the
unit $\mathbb{S}^2$.  First notice that equation
\eqref{UnphysicalGoodEqExplicit} is formally singular at $\tau=\pm 1$
since the coefficient $(\tau^2-1)$ appearing in the principal part
vanishes.  Nonetheless, using the same methods of \cite{Val03a,
  GasVal20} used for the spin-2 equation one can derive an explicit
expression for the exact solution arising from a suitable class of
initial data. This analysis for the wave equation as written in
expression~\eqref{UnphysicalGoodEqExplicit} has already been carried
out in~\cite{MinMacVal22}.  In subsection~\ref{Sol_good_unphys} we
give a brief description of the method and write the solution as
reported in~\cite{MinMacVal22}. Our ultimate goal is to translate this
solution to the physical set up and to compare it with the formal
expansion obtained in~\cite{DuaFenGasHil21} using a method based on
H\"ormander's asymptotic system for the wave equation. In doing so we
will clarify the relation of the logarithmic terms found
in~\cite{DuaFenGasHil21} with those found in solutions to linear
equations propagating in the geometry of the~$F$-cylinder at spatial
infinity as discussed in~\cite{MinMacVal22} for the wave equation and
in~\cite{Val03a, GasVal20} for the spin-1 and spin-2 equations.

%%%%%%%%%%%%%%%%%%%%%%%%%%%%%%%%%%%%%%%%%%%%%%%%%%%%%%%%%%%%%
\subsubsection{Solution in the unphysical picture}
\label{Sol_good_unphys}
%%%%%%%%%%%%%%%%%%%%%%%%%%%%%%%%%%%%%%%%%%%%%%%%%%%%%%%%%%%%%

Following \cite{MinMacVal22} one considers the following Ansatz for
the solution
\begin{align}\label{AnsatzGood}
\phi=
\sum_{p=0}^{\infty}\sum_{\ell=0}^{p}\sum_{m=-\ell}^{m=\ell}
\frac{1}{p!}a_{p;\ell,m}(\tau)Y_{\ell
  m}\rho^{p},
\end{align}
where $Y_{\ell m}$ are the spherical harmonics. Notice that by making
this Ansatz one is implicitly assuming the initial data on
hypersurface
\begin{align}
  \mathcal{S}:=\{ \bmp \in \mathcal{M} \;|\; \tau (\bmp)=0 \},
\end{align}
that is analytic at the cylinder at spatial infinity $\rho=0$, since
\begin{align}\label{ID_good}
\phi|_{\mathcal{S}} =
\sum_{p=0}^{\infty}
\sum_{\ell=0}^{p}\sum_{m=-\ell}^{m=\ell}\frac{1}{p!}a_{p;\ell,m}(0)Y_{\ell
  m}\rho^p, \qquad \dot{\phi}|_{\mathcal{S}} =
\sum_{p=0}^{\infty}\sum_{\ell=0}^{p}\sum_{m=-\ell}^{m=\ell}
\frac{1}{p!}\dot{a}_{p;\ell,m}(0)Y_{\ell
  m}\rho^p,
\end{align}
where $\dot{\phi}|_{\mathcal{S}}:=\partial_\tau \phi|_{\mathcal{S}}$.
Ultimately, the initial data is encoded in the constants
$a_{p;\ell,m}(0)$ and $\dot{a}_{p;\ell,m}(0)$.  Upon substitution of
this Ansatz into equation \eqref{UnphysicalGoodEqExplicit} one obtains
the following ODE for $a_{p;\ell,m}(\tau)$ for each fixed, $\{p,\ell,
m\}$:
\begin{align}\label{ODE_wave_JacobiPoly}
(1-\tau^2)\ddot{a}_{p;\ell,m} +
  2\tau(p-1)\dot{a}_{p,\ell,m}+(\ell+p)(\ell-p+1){a}_{p;\ell,m}=0.
\end{align}
An analysis of this equation given in~\cite{MinMacVal22} gives the
following result
\begin{lemma}[Homogeneous wave equation on the $i^0$-cylinder
    background~\cite{MinMacVal22}]\label{Sol_Good_Jacobi_and_Logs}
  The solution to equation \eqref{ODE_wave_JacobiPoly} is given
  explicitly by:
  \begin{enumerate}[label=(\roman*)]
  \item For $p\geq 1$   and $0\leq \ell \leq p-1$ 
    \begin{align} a(\tau)_{p;\ell,m} =A_{p,\ell,m}
      \bigg(\frac{1-\tau}{2}\bigg)^{p}
      P_{\ell}^{(p,-p)}(\tau) + B_{p,\ell,m}
      \bigg(\frac{1+\tau}{2}\bigg)^{p}P_{\ell}^{(-p,p)}(\tau)
    \end{align}
  \item For  $p\geq 0$   and $\ell=p$:
    \begin{align}\label{Sol_good_highestharmonic}
      {a}_{p;p,m}(\tau) = \bigg(\frac{1-\tau}{2}\bigg)^{p}
      \bigg(\frac{1+\tau}{2}\bigg)^{p}\Bigg(C_{p,p,m} +D_{p,p,m}
      \int_{0}^{\tau} \frac{ds}{(1-s^2)^{p+1}}\Bigg)
    \end{align}
    where~$P^{\alpha,\beta}_{\gamma}(\tau)$ are the Jacobi polynomials
    and~$A_{p,\ell,m}$, $B_{p,\ell,m}$, $C_{p,p,m} $ and $D_{p,p,m} $
    are constants which can be written algebraically in terms of the
    initial data~$a_{p;\ell,m}(0)$ and $\dot{a}_{p;\ell,m}(0)$.
  \end{enumerate}
\end{lemma}
The most interesting feature of solutions obtained through
the~$i^0$-cylinder framework is that even for linear equations such as
the wave equation \eqref{unphysical_good} ---see also \cite{GasVal18,
  Val04d} for the solution to the spin1- and spin-2 equations--- the
expansion close to spatial and null infinity is polyhomogeneous.  To
see this clearly, observe that expanding the integral in
\eqref{Sol_good_highestharmonic} give rise to logarithmic terms.  For
instance for $p=0$ and $p=1$ one has:
\begin{subequations}
\begin{align}
  {a}_{0;0,0}(\tau) & = C_{000} + \tfrac{1}{2} D_{000} (\log(1 + \tau
  )- \log(1 - \tau ))\\ {a}_{1;1,m}(\tau) & = \tfrac{1}{4} (1 - \tau )
  (1 + \tau ) (C_{11m} + \tfrac{1}{4} D_{11m} ( \log(1 + \tau ) -
  \log(1 - \tau ) + 2\tau(1-\tau^2)))
  \end{align}
\end{subequations}
Notice that the solution given by equation \eqref{AnsatzGood} and
Lemma~\ref{Sol_Good_Jacobi_and_Logs} is not an approximate solution:
the sum in~\eqref{AnsatzGood} is an infinite sum and the ODE
\eqref{ODE_wave_JacobiPoly} determining the solution at each order is
solved exactly and explicitly. Moreover, since the modes do not mix,
every partial sum ---from $p=0$ to a fixed finite $p=P$--- constitutes
an exact solution arising from data satisfying
$a_{p;\ell,m}(0)=\dot{a}_{p;\ell,m}(0)=0$ for $p\geq P+1$.

\begin{remark}\label{logfreeRemark}(Log-free initial
  data~\cite{MinMacVal22}).
  \emph{A direct calculation using Lemma
  \ref{Sol_Good_Jacobi_and_Logs} shows that
\begin{align}
C_{p;p,m}= 2^{2p}a_{p;p,m}(0) , \qquad D_{p;p,m}=2^{2p}\dot{a}_{p;p,m}(0)
\end{align}
Therefore, by choosing initial data such that~$\dot{a}_{p;p,m}(0)=0$
one obtains a log-free expansion for~$\phi$.}
\end{remark}

%%%%%%%%%%%%%%%%%%%%%%%%%%%%%%%%%%%%%%%%%%%%%%%%%%%%%%%%%%%%%
\subsubsection{Solution in the physical picture}
\label{Translation-good}
%%%%%%%%%%%%%%%%%%%%%%%%%%%%%%%%%%%%%%%%%%%%%%%%%%%%%%%%%%%%%

It is clear from equation \eqref{Ftophys} that any generic function of
only~$\tau$ or~$\rho$ will lead to expression in the physical
spacetime depending on both~$\tilde{t}$ and~$\tilde{\rho}$ hence,
although the solutions of the Ansatz~\eqref{AnsatzGood} split the
functional form in the $F$-coordinates, this does not translate into a
split in functions depending only on $\tilde{t}$ and $\tilde{\rho}$.
The key observation to understand how the logarithms of
Lemma~\ref{Sol_Good_Jacobi_and_Logs} are expressed in terms of
physical coordinates~$(\tilde{t},\tilde{\rho})$ is the content of the
following Remark.

\begin{remark}
  \emph{ The logarithmic terms in Proposition
  \ref{Sol_Good_Jacobi_and_Logs} appear always in pairs of
  $\log(1-\tau)$ and $\log(1+\tau)$ that can be rewritten simply in
  terms of the boost parameter as $\log \kappa$.  }
  \end{remark}

To see this more clearly, one can write the solution given in
Lemma~\ref{Sol_Good_Jacobi_and_Logs} for the first few orders
explicitly.  Two points of view can be taken: the first one is to
consider ---for generic initial data within the class of
equation~\eqref{ID_good}--- an asymptotic solution close to the
cylinder at spatial infinity up to order~$\mathcal{O}(\rho^{P+1})$:
\begin{align}\label{AnsatzGood_with_error_term}
\phi=
\sum_{p=0}^{P}\sum_{\ell=0}^{p}\sum_{m=-\ell}^{m=\ell}
\frac{1}{p!}a_{p;\ell,m}(\tau)Y_{\ell
  m}\rho^p + \mathcal{O}(\rho^{P+1}).
\end{align}
The second one consists in exploiting the fact that the solution given
in~\ref{Sol_Good_Jacobi_and_Logs} is exact, for the class of initial
data \eqref{ID_good}. Hence, in order to have a solution written as a
finite sum of terms, one can simply restrict the initial data so that
the sum in~\eqref{AnsatzGood} ends at a finite~$p=P$. For instance
consider the solution arising from initial data satisfying
\begin{align}
a_{p;\ell,m}(0)=\dot{a}_{p;\ell,m}(0)=0 \qquad \text{for} \qquad p\geq
2.
\end{align}
The (exact) solution for the unphysical field~$\phi_{g}$ simply reads
\begin{align}
\phi = a_{0,0,0}(\tau)Y_{00}+ \left[a_{1,0,0}(\tau) Y_{00} +
  a_{1,1,-1}(\tau) Y_{1,-1} + a_{1,1,0}(\tau)\right] \rho
\end{align}
Substituting~$a_{p,\ell,m}$ using Lemma~\ref{Sol_Good_Jacobi_and_Logs}
and writing the terms using the definition for~$\kappa$, in
expression~\eqref{boostphysical}, leads to
\begin{align}
\phi &= \frac{1}{2} (2 C_{000} + D_{000} \log\kappa ) Y_{00} +
\frac{1}{4} \rho \Big(2 Y_{00} ( A_{100} (1 - \tau ) + B_{100} (1 +
\tau )\Big)  \nonumber\\
&\quad+ Y_{1-1} (1 - \tau ) (1 + \tau ) \bigg(C_{11-1}
+ \frac{1}{4} D_{11-1} \bigg(\log \kappa + \frac{2 \tau }{1 - \tau
  ^2}\bigg)\bigg)  \nonumber\\
&\quad+ Y_{10} (1 - \tau ) (1 + \tau )
\bigg(C_{110} + \frac{1}{4} D_{110} \bigg(\log \kappa + \frac{2 \tau
}{1 - \tau ^2}\bigg)\bigg)  \nonumber\\
&\quad+ Y_{11} (1 - \tau ) (1 + \tau
) \Big(C_{111} + \frac{1}{4} D_{111} \Big(\log \kappa  + \frac{2 \tau }{1 -
  \tau ^2}\Big)\Big),\label{Unphysical_good_sol_order1}
\end{align}
where the constants $C_{p,\ell,m}$, $D_{p,\ell,m}$, $A_{p,\ell,m}$
and~$B_{p,\ell,m}$ are determined by the non-trivial initial
data~$a_{p;\ell,m}(0)$ and $\dot{a}_{p;\ell,m}(0)$ for $0\leq p\leq
1$, $0 \leq \ell\leq p$ and $-\ell \leq m \leq \ell$.  Hence,
recalling the relation between the physical and unphysical good
fields~$\tilde{\phi}=\Theta \phi$, using equation \eqref{CF-theta} for
the conformal factor, and writing
equation~\eqref{Unphysical_good_sol_order1} expressed through the
physical advanced and retarded times one gets
\begin{align}\label{PhysicalFieldSolution}
  \tilde{\phi} &=  \frac{C_{000} Y_{00}}{\tilde{\rho}} +
  \frac{Y_{00}}{2 \tilde{\rho}}\Big( \frac{A_{100}}{ \tilde{v}} -
  \frac{B_{100}}{ \tilde{u} }+ D_{000}\log\kappa \Big) +
  \frac{C_{11-1} Y_{1-1}+ C_{111} Y_{11} + C_{110} Y_{10}}{4
    \tilde{\rho}^2} \nonumber \\
  & \quad+ \frac{1}{32
    \tilde{\rho}^2}
  \Big(\frac{\tilde{u}}{\tilde{v}}-\frac{\tilde{v}}{\tilde{u}}+
  2\log \kappa\Big) (D_{11-1}Y_{1-1} + D_{110}Y_{10} +D_{111}Y_{11 }).
\end{align}
recalling that the boost parameter~$\kappa$ can be written in terms of
physical coordinates as~$\kappa=-\tilde{v}/\tilde{u}$ ---see
Proposition~\ref{Prop:FframeToPhsyicalframe}--- one obtains an
explicit exact solution for the physical field and written in physical
coordinates for the type of initial data considered.  Repeating the
above explicit calculation, following the point of view that general
initial data in the class of equation~\eqref{ID_good} is considered,
and keeping the error order term in
equation~\eqref{AnsatzGood_with_error_term} one obtains the following:

\begin{proposition}
  \label{prop:good}
  \emph{ The solution $\tilde{\phi}_g$ to equation \eqref{gbu-good}
  ---the physical \emph{good} equation--- arising from analytic
  initial data close to the cylinder at spatial infinity $I$ has the
  following formal expansion
  \begin{flalign}\label{Sol_Good_From_Cylinder}
   \tilde{\phi}_g = & \frac{C_{000} Y_{00}}{\tilde{\rho}} +
   \frac{Y_{00}}{2 \tilde{\rho}}\Big( \frac{A_{100}}{
     \tilde{v}}-\frac{B_{100}}{ \tilde{u} } + D_{000}\log\kappa \Big)
   + \frac{C_{11-1} Y_{1-1}+ C_{111} Y_{11} + C_{110} Y_{10}}{4
     \tilde{\rho}^2} \nonumber \\ & + \frac{1}{32
     \tilde{\rho}^2}
   \Big(\frac{\tilde{u}}{\tilde{v}}-\frac{\tilde{v}}{\tilde{u}}+
   2\log \kappa\Big) (D_{11-1}Y_{1-1} + D_{110}Y_{10} +D_{111}Y_{11 })
   + \mathcal{O}(\tilde{\rho}^{-3}) .
  \end{flalign}
  }
\end{proposition}

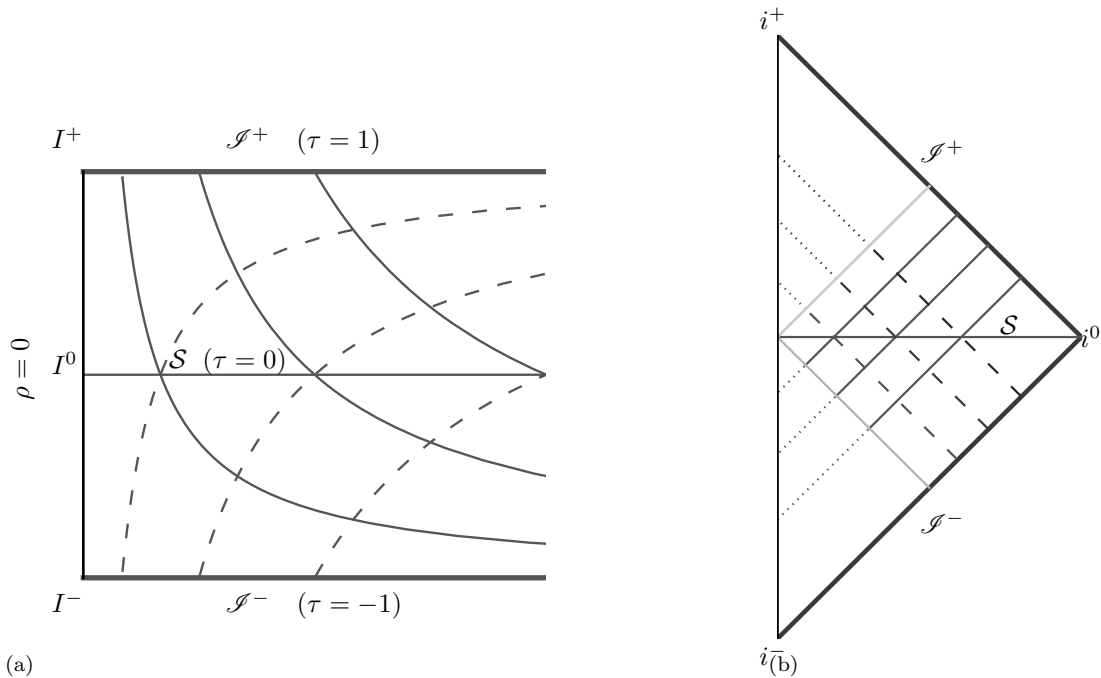
\begin{figure}[t!]
  \begin{subfigure}[]{}
    \includegraphics[width=7cm]{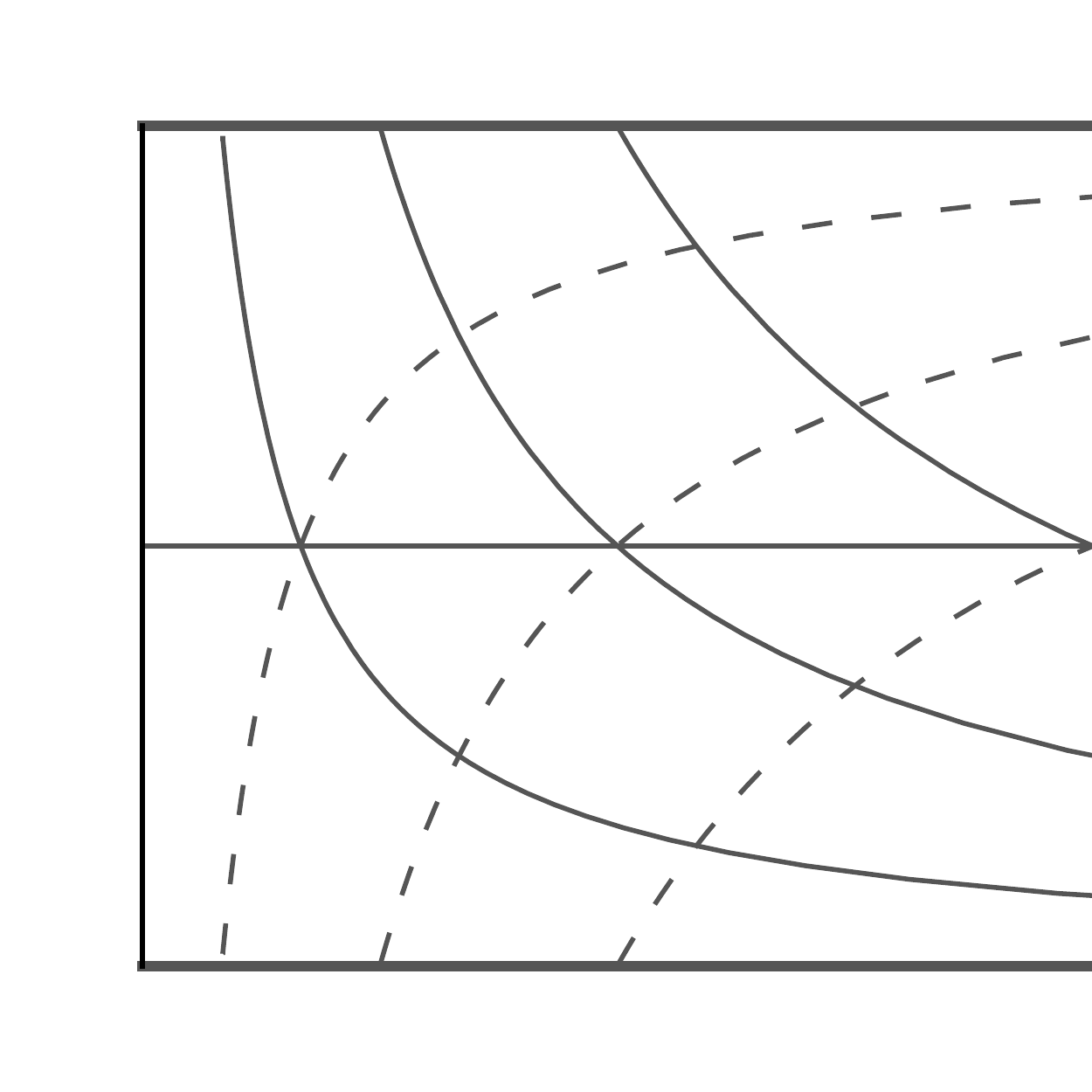}
    \put(-120,10){$\mathscr{I}^{-}\quad (\tau=-1)$}
    \put(-185,10){$I^{-}$} \put(-120,185){$\mathscr{I}^{+}\quad (\tau=1)$}
    \put(-185,185){$I^{+}$} \put(-185,100){$I^{0}$}
    \put(-141,102){$\mathcal{S} \;\; (\tau=0)$}
    \put(-200,90){\rotatebox{90}{$\rho=0$} }
    %%%%%%%%%%%%%%%%%%%%%%%%%%%%%
    \put(140,40){$\mathscr{I}^{-}$} \put(140,180){$\mathscr{I}^{+}$}
    \put(200,110){$i^{0}$} \put(80,230){$i^{+}$} \put(80,-10){$i^{-}$}
    \put(170,115){$\mathcal{S}$} \qquad\qquad\qquad \qquad
  \end{subfigure}
  \begin{subfigure}[]{}
    \includegraphics[width=4cm]{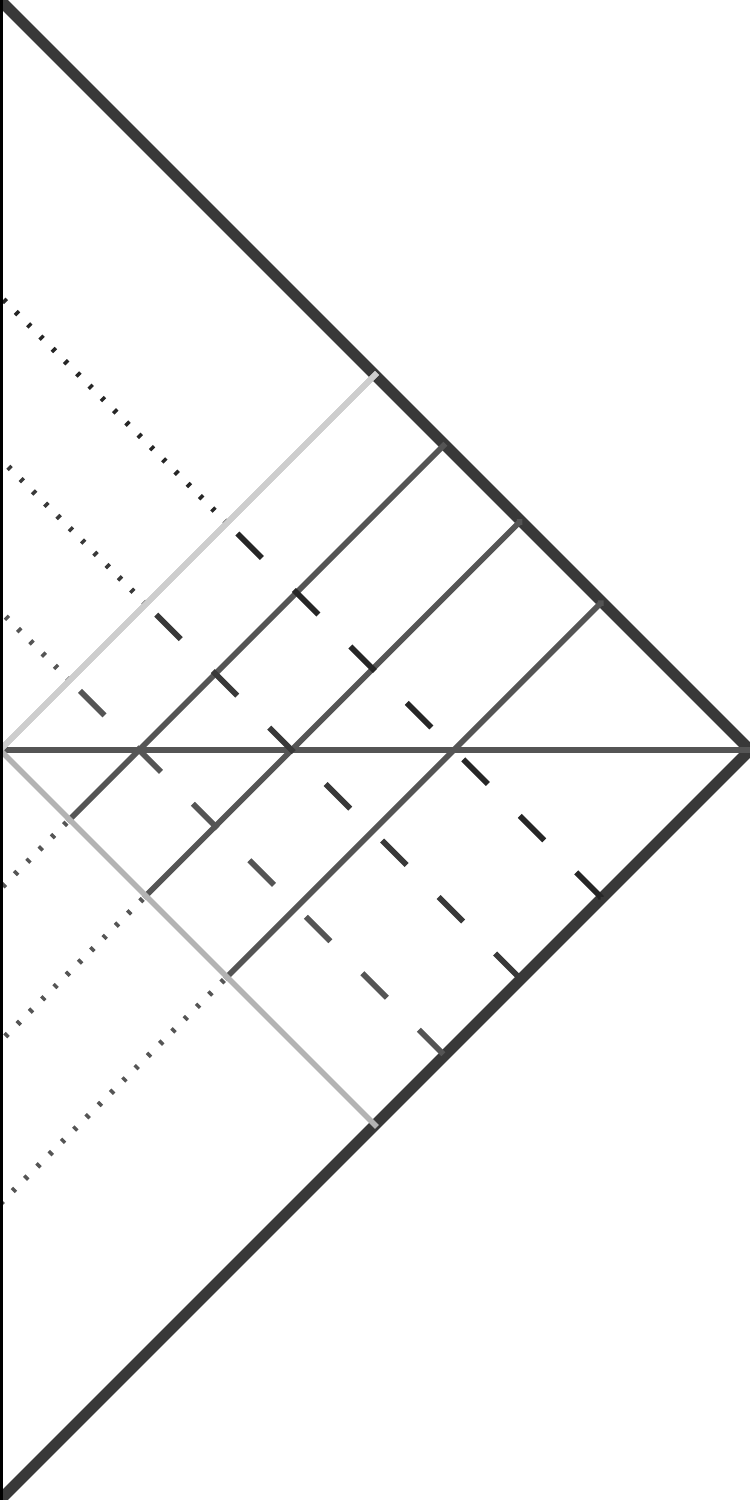}
  \end{subfigure}  
      \caption{Panel (a) shows a coordinate diagram with the cylinder
        at spatial infinity for the Minkowski spacetime. Panel (b)
        shows the Penrose diagram of the Minkowski spacetime with the
        lightcone at the origin is depicted in light-gray: the region
        where the~$F$-coordinates are valid is the complement of this
        cone (neighbourhood of~$i^0$).  The continuous lines represent
        null surfaces~$\tilde{u}=-|\tilde{u}_\star|$ the dashed ones
        represent null surfaces~$\tilde{v}=|\tilde{v}_{\star}|$. }
       \label{fig:CylinderAndPenroseDiagram}
\end{figure}

\begin{remark}
   \emph{The terms with~$\log \kappa$ are real valued functions since
   the range of validity of the coordinates corresponds to the
   complement of the lightcone at the origin of the Minkowski
   spacetime and hence~$\tilde{u}<0$ and~$\tilde{v}>0$ so
   that~$\kappa>0$ ---see Figure~\ref{fig:CylinderAndPenroseDiagram}.}
\end{remark}

%%%%%%%%%%%%%%%%%%%%%%%%%%%%%%%%%%%%%%%%%%%%%%%%%%%%%%%%%%%%%
%%%%%%%%%%%%%%%%%%%%%%%%%%%%%%%%%%%%%%%%%%%%%%%%%%%%%%%%%%%%%

Exploiting Remark~\ref{logfreeRemark} it is clear how to identify
initial data for the physical good field which leads to log-free
expansions. Splitting the initial data for~$\dot{\phi}$ as
\begin{align}
  \dot{\phi}|_{\mathcal{S}} =\sum_{p=0}^{\infty}
  \sum_{m=-p}^{m=p}\frac{1}{p!}\dot{a}_{p;p,m}(0)Y_{\ell m}\rho^p
  + \sum_{p=1}^{\infty}\;\;\sum_{\ell=1}^{\max\{1, \; p-1\}}
  \sum_{m=-\ell}^{m=\ell}\frac{1}{p!}\dot{a}_{p;\ell,m}(0)Y_{\ell m}\rho^p ,
\end{align}
then it follows that to have a \emph{log-free expansion at all orders}
one needs special initial data of the form
\begin{align}
\dot{\phi}|_{\mathcal{S}} =
\sum_{p=1}^{\infty}\;\;\sum_{\ell=1}^{{\max\{1, \; p-1\}}}\sum_{m=-\ell}^{m=\ell}
\frac{1}{p!}\dot{a}_{p;\ell,m}(0)Y_{\ell m}\rho^p.
\end{align}
\begin{remark}\label{remarkID}
  \emph{Notice also that the appearance of the logs is irrespective of
  the value of~$a_{p,\ell,m}(0)$ and hence of the initial data
  for~$\phi|_\mathcal{S}$ as given in equation~\eqref{ID_good}. Using
  Proposition~\ref{Prop:FframeToPhsyicalframe} one has
  that~$\partial_\tau \phi|_{\mathcal{S}} =
  \tilde{\rho}\partial_{\tilde{t}}\phi|_{\mathcal{S}}$. Hence,
  recalling that $\phi=\Theta^{-1} \tilde{\phi}$, then $
  \dot{\phi}|_{\mathcal{S}} =
  \tilde{\rho}^2\partial_{\tilde{t}}\tilde{\phi}|_{\mathcal{S}}$.  }
  \end{remark}
 This leads to the following
\begin{proposition}\label{logfree_data_physgood}
  \emph{Initial data for the physical good field satisfying
\begin{align}
\tilde{\phi}|_{\mathcal{S}} =
\sum_{p=0}^{\infty}\sum_{\ell=0}^{p}\sum_{m=-\ell}^{m=\ell}
\frac{1}{p!}a_{p;\ell,m}(0)Y_{\ell
  m}\tilde{\rho}^{p-1}, \quad
\partial_{\tilde{t}}\tilde{\phi}|_{\mathcal{S}}=
\sum_{p=1}^{\infty}\sum_{\ell=1}^{{\max\{1, \; p-1\}}}
\sum_{m=-\ell}^{m=\ell}\frac{1}{p!}\dot{a}_{p;\ell,m}(0)Y_{\ell m}
\tilde{\rho}^{-p-2}.
\end{align}
gives rise to a log-free expansion close to $I$.}
\end{proposition}
If one is only interested in suppressing the \emph{leading logarithm}
---that associated to~$D_{000}$--- then requiring that the initial
data for the physical good field satisfies
\begin{align}\label{naive_conditionID}
  \partial_{\tilde{t}}\tilde{\phi}|_{\mathcal{S}}=
  \mathcal{O}(\tilde{\rho}^{-3})
\end{align}
ensures that~$D_{000}=0$. Later it will be shown that the
condition~\eqref{naive_conditionID}, is sufficient but not necessary
to do so.  Moreover, it does not prevent the appearance of higher order
logs.

%%%%%%%%%%%%%%%%%%%%%%%%%%%%%%%%%%%%%%%%%%%%%%%%%%%%%%%%%%%%%
\subsection{The ugly equation}
%%%%%%%%%%%%%%%%%%%%%%%%%%%%%%%%%%%%%%%%%%%%%%%%%%%%%%%%%%%%%

It was observed in~\cite{DuaFenGasHil21} that the simple linear wave
equation~\ref{gbu-ugly}, motivated by the equation of motion for
certain components of the physical metric in generalised harmonic
gauge, gives rise to polyhomogeneous expansions near null
infinity. Using the
identity~\eqref{General_Wave_Conformal_Transformation} and the
specific feature of the~$i^0$-cylinder background discussed in
Section~\ref{TheCylinderAtSpatialInfinity}, one gets

\begin{align}
  \square \phi & =
  \frac{2}{\tilde{\rho}}\Theta^{-3} \tilde{\nabla}_{\tilde{t}}\tilde{\phi}
  \nonumber\\ & =
  2\Theta^{-1} \tilde{\nabla}_{\tilde{t}}(\Theta^{-1}\tilde{\phi})
  \nonumber\\ & =
  ( \kappa \nabla _{\uell}\phi + \kappa^{-1}\nabla _{\ell}\phi)
\end{align}

where from the first to the second line equation \eqref{CF-theta} was
employed.
\begin{remark}\emph{Although it may appear obvious that the conformal
  factor~$\Theta$ is a function of the physical radial
  coordinate~$\tilde{\rho}$ only, this fact is not assumed a priori in
  the general framework of the conformal Einstein field equations.  In
  the case of the Minkowski spacetime the relation between the
  physical and unphysical coordinates is explicitly known so that this
  relation be given in closed form.}
\end{remark}
Observe as well that although the \emph{unphysical ugly} equation
\begin{align}\label{unphysical_ugly_equation}
  \square \phi & =   \kappa \nabla _{\uell}\phi +
  \kappa^{-1}\nabla _{\ell}\phi
\end{align}
does not contain any~$\Theta^{-1}$ term, the boost parameter~$\kappa$
and its inverse appear in equation \eqref{unphysical_ugly_equation}
which is singular at~$\tau=\pm 1$ respectively. The unphysical ugly
equation in the~$F$-coordinates reads
\begin{align}
  ( \tau ^2-1) \partial _{\tau}^2 \phi -2 \rho \tau
  \partial_{\tau}\partial _{\rho}\phi + \rho ^2 \partial _{\rho}^2\phi
  + 2 (\tau +q) \partial _{\tau}\phi + 4  \rho \tau ( \tau ^2-1)^{-1}
  \partial _{\rho}\phi + \Delta _{\mathbb{S}^{2}{}}{}\phi =0.
\end{align}
Using the Ansatz of equation~\eqref{AnsatzGood} then renders the
following ODE for~$a_{p;\ell,m}$
\begin{align}\label{ODE_ugly}
(1-\tau^2)\ddot{a}_{p;\ell,m} +
  2(\tau(p-1)-1)\dot{a}_{p,\ell,m}+\Big[(\ell+p)(\ell-p+1)
    +\frac{p\tau}{1-\tau^2}\Big]{a}_{p;\ell,m}=0.
\end{align}
Similarly to equation~\eqref{UnphysicalGoodEqExplicit}, this equation
is formally singular at~$\tau=\pm1$, hence to contrast
equations~\eqref{UnphysicalGoodEqExplicit} and~\eqref{ODE_ugly}, the
latter can be written more fairly if we multiply through by~$1-\tau^2$
\begin{align}\label{ODE_ugly_fairlywritten}
(1-\tau^2)^2\ddot{a}_{p;\ell,m} +
  2(1-\tau^2)(\tau(p-1)-1)\dot{a}_{p,\ell,m}+\Big[(\ell+p)(\ell-p+1)(1-\tau^2)
    +p\tau \Big]{a}_{p;\ell,m}=0.
\end{align}
Unfortunately, equation~\eqref{ODE_ugly_fairlywritten} does not have
the form of a Jacobi equation in which the theory of special functions
of~\cite{Sze78} can be applied.
Presumably,~\eqref{ODE_ugly_fairlywritten} lies in the class of the
Heun equation and could be solved accordingly in terms of special
functions and power series. Although the latter described strategy
could be pursued, an alternative, cleaner, approach is to realise that
the ugly model equation can be written as a good equation for a
suitably defined field and then apply the analysis of
subsection~\ref{Subsection:GoodEquation} on this newly defined field.
To do so it is convenient to rewrite it in terms of null directions.
The physical ugly equation~\eqref{gbu-ugly} expressed using the
physical outgoing and incoming null vectors reads
\begin{align}\label{classical_ugly_phys_LLbar}
\tilde{\square}\tilde{\phi} - \frac{1}{\tilde{\rho}}(L\tilde{\phi}
+ \underline{L}\tilde{\phi})=0.
\end{align}
Using that~$\tilde{\square}$ can be written in terms of~$L$
and~$\underline{L}$ as
\begin{align}\label{PhysWaveOp}
  \tilde{\square}\tilde{\phi}:=
  -\frac{1}{\tilde{\rho}}L\underline{L}(\tilde{\rho}\tilde{\phi})
+ \frac{1}{\tilde{\rho^2}} \Delta _{\mathbb{S}^{2}} \tilde{\phi},
\end{align}
and using the expressions for~$L$ and~$\underline{L}$ as given in
equation~\eqref{PhysicalFrameExplicit}, the following `commutation
relation' can be derived
\begin{align}\label{CommutationLbar1}
  \underline{L}(\tilde{\rho}^2\tilde{\square}\tilde{\phi})
  -\tilde{\rho}\tilde{\square}(\tilde{\rho}\underline{L} \tilde{\phi})
  =
  \underline{L}(\tilde{\rho}(L+ \underline{L})\tilde{\phi}).
\end{align}
The latter expression can be rewritten as
\begin{align}\label{commutation_rewrriten}
\underline{L}\Bigg(\tilde{\rho}^2\Big( \tilde{\square} \tilde{\phi} -
\frac{1}{\tilde{\rho}}(L\tilde{\phi} + \underline{L}\tilde{\phi})\Big)
\Bigg)
=\tilde{\rho}\tilde{\square}(\tilde{\rho}\underline{L}\tilde{\phi}).
\end{align}
Hence, the physical ugly equation~\eqref{classical_ugly_phys_LLbar}
implies that
\begin{align}\label{UglyAsAGood}
\tilde{\square}\tilde{\Phi}=0.
\end{align}
where
\begin{align}\label{DefUglyAsAGood}
\tilde{\Phi} = \tilde{\rho}\underline{L}\tilde{\phi}.
\end{align}
The initial data for the auxiliary problem~\eqref{UglyAsAGood} is not
free if the aim is to construct a solution to the physical ugly
equation~\eqref{classical_ugly_phys_LLbar}. To see this, notice that
from equation~\eqref{commutation_rewrriten} it follows that by solving
the \emph{auxiliary problem} \eqref{UglyAsAGood} and
\eqref{DefUglyAsAGood}, one is not necessarily obtaining a solution to
the ugly equation~\eqref{classical_ugly_phys_LLbar}, but rather to the
more general equation
\begin{align}\label{gnraleqwithQ}
\tilde{\square} \tilde{\phi} - \frac{1}{\tilde{\rho}}(L\tilde{\phi} +
\underline{L}\tilde{\phi})= Q,
\end{align}
where~$Q$ is smooth function of the physical
coordinates~$\tilde{x}^\mu$ that
satisfies~$\underline{L}(\tilde{\rho}^2Q)=0$.  The source term~$Q$
encodes the relation between the data for the physical ugly
field~$\tilde{\phi}$ and the \emph{auxiliary physical good
field}~$\tilde{\Phi}$.  To clarify this relation observe that using
equation \eqref{DefUglyAsAGood} and writing
equation~\eqref{gnraleqwithQ} as
\begin{align}
  -\partial_{\tilde{t}}^2\tilde{\phi} + \Delta \tilde{\phi}
  + \frac{2}{\tilde{\rho}}\partial_{\tilde{t}}\tilde{\phi} = Q,
\end{align}
where~$\Delta$ denotes the Laplace operator of 3-dimensional Euclidean
space, it follows that the initial data is related via
\begin{subequations}
  \begin{align}
    \tilde{\Phi}|_{\mathcal{S}} & =
    \left[\tilde{\rho}(\partial_{\tilde{t}}-\partial_{\tilde{\rho}})
    \tilde{\phi}\right]_{\mathcal{S}},\\ 
    \partial_{\tilde{t}}\tilde{\Phi}|_{\mathcal{S}} &=
    \left[\tilde{\rho}(\partial^{2}_{\tilde{t}}
    -\partial_{\tilde{\rho}}\partial_{\tilde{t}})\tilde{\phi}\right]_{\mathcal{S}}
    = \left[\tilde{\rho}(\Delta \tilde{\phi} + 2
    \tilde{\rho}^{-1}\partial_{\tilde{t}}\tilde{\phi}-Q
    -\partial_{\tilde{\rho}}\partial_{\tilde{t}}\tilde{\phi})\right]_{\mathcal{S}}
  \end{align}
\end{subequations}
Once the appropriate initial data for the auxiliary
field~$\tilde{\Phi}$ has been obtained, then using
that~$\underline{L}=2\partial_{\tilde{u}}$, integrating the
expression~\eqref{DefUglyAsAGood} along the incoming null geodesic,
one gets the formal expression
\begin{align}\label{Int_ugly_from_good}
\tilde{\phi}=
\frac{1}{2}
\int_{\tilde{u}_{\star}}^{\tilde{u}}\frac{\tilde{\Phi}(\bar{u},\tilde{v})
} {\tilde{\rho}(\bar{u},\tilde{v})}d\bar{u}.
\end{align}
Hence, the analysis for the physical good equation given in
subsection~\ref{Translation-good} and summarised in
Proposition~\ref{prop:good} provides a general solution to the
auxiliary problem ---modulo adjusting the initial data as describe
before.  Multiplying by~$\/\tilde{\rho}$ and integrating the
asymptotic expansion for the physical good field given in
Lemma~\ref{prop:good}, one obtains
\begin{flalign}
  \tilde{\phi} = & \int_{\tilde{u}_{\star}}^{\tilde{u}}
  \Big\{\frac{C_{000}Y_{00}}{\tilde{\rho}^2} +
  \frac{Y_{00}}{2 \tilde{\rho}^2}\Big(
  \frac{A_{100}}{ \tilde{v}}-\frac{B_{100}}{ \bar{u} }
  + D_{000}\log\kappa \Big) +
  \frac{C_{11-1} Y_{1-1}+ C_{111} Y_{11} + C_{110} Y_{10}}{4
    \tilde{\rho}^3} \nonumber \\
  & + \frac{1}{32
    \tilde{\rho}^3}\Big(\frac{\bar{u}}{\tilde{v}}
  -\frac{\tilde{v}}{\bar{u}}+
  2\log \kappa\Big) (D_{11-1}Y_{1-1} + D_{110}Y_{10} +D_{111}Y_{11 })
  + \mathcal{O}(\tilde{\rho}^{-4}) \Big\} d\bar{u}. \nonumber
\end{flalign}
which, after integration renders,
\begin{flalign}\label{Sol_Ugly_From_Cylinder_RAW}
  \tilde{\phi} = & \frac{2C_{000} Y_{00}}{\tilde{\rho}} +
  \frac{Y_{00}(A_{100}-B_{100})}{\tilde{v}\tilde{\rho}}-
  \frac{2B_{100}Y_{00}}{\tilde{v}^2}\log\Big(-\frac{\tilde{u}}
       {2\tilde{\rho}}\Big)+
  \frac{Y_{00}D_{000}\log \kappa}{\tilde{\rho}} \nonumber \\ &
  + \frac{2D_{000}Y_{00}}{\tilde{v}}
  \log\Big(-\frac{\tilde{u}}{2\tilde{\rho}}\Big) + \frac{C_{11-1}
    Y_{1-1}+ C_{111} Y_{11} + C_{110} Y_{10}}{4 \tilde{\rho}^2} 
  - \frac{D_{11-1}Y_{1-1} + D_{110}Y_{10} +D_{111}Y_{11 }}
  {8\tilde{v}\tilde{\rho}} \nonumber \\ &
  +  \frac{D_{11-1}Y_{1-1} + D_{110}Y_{10} +D_{111}Y_{11 }}{16
    \tilde{\rho}^2}\log \kappa + \mathcal{O}(\tilde{\rho}^{-3}).
 \end{flalign}
 \begin{remark}
  \emph{The term~$\log(-\frac{\tilde{u}}{2\tilde{\rho}})$ is a real
  valued function since the range of validity of the inversion
  unphysical coordinates~$x^\mu$ in
  equation~\eqref{UnphysicalCartesianToPhysicalCartesian} ---used in
  turn to build the~$F$-coordinates--- correspond to the region
  determined by~$\tilde{\rho}>0$ with $\tilde{u}<0$
  and~$\tilde{v}>0$.}
 \end{remark}
 \begin{remark}
  \label{Remark:ugly}
  \emph{Observe that there are two types of logarithmic terms
  appearing in the expression \eqref{Sol_Ugly_From_Cylinder_RAW},
  those coming from integration of terms such
  as~$\tilde{\rho}^{-1}\tilde{u}^{-1}$ and those inherited from the
  terms containing~$\log \kappa$ in the good field.  The logs reported
  in~\cite{DuaFenGasHil21} can only correspond to the former since in
  the last reference the good field does not contain any logarithmic
  term.}
\end{remark}
Choosing initial data for the auxiliary good field ensuring $Q=0$
within the working Ansatz~\eqref{AnsatzGood} boils down to setting to
zero some of the constants in equation~\eqref{ID_good}. To the order
shown in equation~\eqref{Sol_Ugly_From_Cylinder_RAW} this corresponds
to setting~$A_{100}=D_{000}=D_{110}=D_{111}=D_{11-1}=0$ which in
particular gets rid of all the logarithmic terms directly inherited
from the good field. The last discussion is summarised in the
following
\begin{proposition}
  \label{prop:ugly}
  \emph{Let~$\tilde{\phi}_u$ be a solution to
  equation~\eqref{gbu-ugly} constructed from solving
  equations~\eqref{UglyAsAGood} and~\eqref{DefUglyAsAGood} with
  analytic initial data close to~$I$.  Then, the
  field~$\tilde{\phi}_u$ has the following asymptotic expansion
  near~$I$:
\begin{flalign}\label{Sol_Ugly_From_Cylinder}
  \tilde{\phi}_u = & \frac{2C_{000} Y_{00}}{\tilde{\rho}} 
  -\frac{Y_{00}B_{100}}{\tilde{v}\tilde{\rho}}
  -\frac{2B_{100}Y_{00}}{\tilde{v}^2}\log
  \Big(-\frac{\tilde{u}}{2\tilde{\rho}}\Big) 
   + \frac{C_{11-1}
    Y_{1-1}+ C_{111} Y_{11} + C_{110} Y_{10}}{4 \tilde{\rho}^2} 
  +  \mathcal{O}(\tilde{\rho}^{-3}).
\end{flalign}
  }
\end{proposition}

%%%%%%%%%%%%%%%%%%%%%%%%%%%%%%%%%%%%%%%%%%%%%%%%%%%%%%%%%%%%%
\subsection{The bad equation}
%%%%%%%%%%%%%%%%%%%%%%%%%%%%%%%%%%%%%%%%%%%%%%%%%%%%%%%%%%%%%

Using the identity~\eqref{General_Wave_Conformal_Transformation} and
substituting the physical bad equation \eqref{gbu-bad}, one gets
\begin{align}
  \square \phi_b & =
  \Theta^{-3}(\tilde{\nabla}_{\tilde{t}}\tilde{\phi}_g)^2 , \nonumber\\ &
  = \Theta^{-1}(\tilde{\nabla}_{\tilde{t}}\phi_g)^2, \nonumber\\ & =
  \frac{1}{4} \Theta ( \kappa \nabla _{\uell}\phi_g + \kappa^{-1}\nabla
  _{\ell}\phi_g)^2,
\end{align}
where from the first to the second line equation \eqref{CF-theta} was
employed. From the second to the third line, the relation between the
physical and unphysical fields~$\tilde{\phi}=\Theta \phi$ and the
results of Proposition \ref{Prop:FframeToPhsyicalframe} were
used. Notice that, similarly to the case of the ugly equation,
although there are no singular terms of the form~$\Theta^{-1}$, the
\emph{unphysical bad equation}
\begin{align}\label{unphysical_bad}
  \square \phi_b = \frac{1}{4}\Theta ( \kappa \nabla _{\uell}\phi_g +
  \kappa^{-1}\nabla _{\ell}\phi_g)^2,
\end{align}
does contain terms which will be singular at~$\mathscr{I}^{+}$
and~$\mathscr{I}^{-}$, due to the presence of the boost parameter in
the form $\Theta\kappa^2$ and $\Theta\kappa^{-2}$, respectively.
Given that the good and bad fields are decoupled, the analysis of the
bad equation follows as a sub-case of the analysis of the following
wave equation with sources
\begin{align}
  \square \phi = f
\end{align}
where~$f=f(\tau,\rho,\theta^A)$. Proceeding as before, using the
Ansatz~\eqref{AnsatzGood}, the calculation boils down to analysing the
ODE:
\begin{align}\label{ODE_wave_JacobiPoly_source}
(1-\tau^2)\ddot{a}_{p;\ell,m} +
  2\tau(p-1)\dot{a}_{p,\ell,m}+(\ell+p)(\ell-p+1){a}_{p;\ell,m}
  =f_{p;\ell,m}(\tau).
\end{align}
where~$f_{p;\ell,m}(\tau)$ arises from expanding $f$ according to the
Ansatz~\eqref{AnsatzGood}. The analysis of this wave equation has been
given in Appendix D of~\cite{MinMacVal22} which we recall here:

\begin{lemma}[Inhomogeneous wave equation on the $i^0$-cylinder
    background~\cite{MinMacVal22}]\label{ODE_bad_sol} The solution
  $a_{p;\ell,m}(\tau)$ of equation \eqref{ODE_wave_JacobiPoly_source}
  can be written as:
\begin{align}
  a_{p;\ell,m}(\tau)=a^{H}_{1:p;\ell,m}(\tau)b_{1:p;\ell,m}(\tau) +
  a^{H}_{2:p;\ell,m}(\tau)b_{2:p;\ell,m}(\tau),
\end{align}
where~$a^{H}_{1:p;\ell,m}(\tau)$ and $a^{H}_{2:p,\ell,m}(\tau)$ are
two independent solutions to the homogeneous problem ---namely,
with~$f(\tau)=0$--- while~$b_{1:p;\ell,m}(\tau)$ and
$b_{2:p;\ell,m}(\tau)$ are given by
\begin{subequations}
\begin{align}
 & b_{1:p;\ell,m}(\tau)= F_{p;\ell,m}
  -  \int_{0}^{\tau}\frac{a_{2:p;\ell,m}(s)f_{p,\ell,m}(s)}{W_\star(1-s^2)^{p}}ds,
  \\ & b_{2:p,\ell,m}(\tau)= G_{p;\ell,m} -
  \int_{0}^{\tau}\frac{a_{1:p;\ell,m}(s)f_{p,\ell,m}(s)}{W_\star(1-s^2)^{p}}ds.
\end{align}
\end{subequations}
where~$F_{p;\ell,m}$, $G_{p;\ell,m}$ and~$W_\star$ are constants.
\end{lemma}

Naturally, the behaviour of the solution $a(\tau)$ hence depends on
the regularity of the source $f(\tau)$.  In the case of the bad
equation~\eqref{unphysical_bad} the source term is given by
\begin{align}
 f=\frac{1}{4}\Theta ( \kappa \nabla _{\uell}\phi_g +
 \kappa^{-1}\nabla _{\ell}\phi_g)^2
\end{align}
where~$\phi_g$ is the solution to equation~\eqref{unphysical_good},
hence, in principle, $f(\tau)$ could contain both poles at $\tau=\pm
1$ and logarithmic terms of the form $\log(1\pm\tau)$.  More
explicitly, the source reads
\begin{align}
  f= -\frac{\rho}{1 - \tau ^2} ((1 - \tau ^2) \partial _{\tau}\phi_g
  +2 \rho \tau
  \partial _{\rho}\phi_g )^2.
\end{align}
A first observation is that the logarithmic terms coming from~$\phi_g$
do not give rise to logarithmic terms in the source~$f$ since
\begin{align}
  \partial_{\tilde{t}}\log \kappa = \frac{1}{2}(\partial_{\tilde{u}}
  +\partial_{\tilde{v}})\log \kappa =-\frac{1}{2}
  \Big(\frac{1}{\tilde{u}}-\frac{1}{\tilde{v}}\Big).
\end{align}
A direct calculation using the solutions for the unphysical good
field~$\phi_g$ as determined by Lemma \ref{Sol_Good_Jacobi_and_Logs}
gives:
\begin{align}\label{sourceexplicit1}
  f= & \frac{\rho D_{000}{}^2 Y_{00}{}^2 }{1 - \tau ^2}
  + \frac{\rho^2 D_{000}  Y_{00}}{2(1-\tau^2)}
  \Big( D_{11-1} Y_{1-1} +  D_{110} Y_{10} +  D_{111}
  Y_{11} - 2 A_{100} Y_{00} (1 - \tau )^2  \nonumber \\
  & +2 B_{100} Y_{00} (1 +  \tau )^2\Big)
  + \mathcal{O}(\tilde{\rho}^{-3}).
\end{align}
In general,~$f$ will contain products of spherical harmonics, hence,
to extract~$f_{p;\ell,m}(\tau)$ one would need to
express~$Y_{\ell,m}Y_{{\ell'},{m'}}$ in terms of linear combinations
of~$Y_{L,M}$ where $|\ell-\ell'| \leq L \leq \ell+\ell'$ and $M=m+m'$,
via the Clebsch-Gordan coefficients. Fortunately, to the pursued
order, this will not be necessary since one of the factors
is~$Y_{00}$, which is constant.  Using equation
\eqref{sourceexplicit1}, along with
Lemmas~\ref{Sol_Good_Jacobi_and_Logs} and~\ref{ODE_bad_sol}, gives the
following expansion for the unphysical bad field~$\phi_b$
\begin{align}\label{Bad_unphysical}
  \phi_b = C_{000} F_{000} Y_{00} + \tfrac{1}{2} G_{000} \log(\kappa )
  Y_{00} + \rho (H_1 + H_2\tau +H_3\tau^2+\log \kappa (H_4 +
  H_5\tau^2)) + \mathcal{O}(\rho^2),
\end{align}
where,
\begin{align}
  H_1 & =\frac{1}{2} (A_{100} F_{100} + B_{100} G_{100}) Y_{00} +
  \frac{1}{4}( C_{11-1} F_{11-1} Y_{1-1} + C_{110} F_{110} Y_{10} +
  C_{111} F_{111} Y_{11}), \nonumber \\
  H_2 &=-\frac{Y_{00}}{8\pi^{1/2}
    W_{*}{}} \Big(A_{100} (B_{100} D_{000}{}^2 + 4 \pi^{1/2} W_{*}{}
  F_{100}) -4 \pi^{1/2} W_{*}{} B_{100} G_{100}\Big) \nonumber \\ &
  \quad + \frac{1}{8}(G_{11-1} Y_{1-1} + G_{110} Y_{10} + G_{111}
  Y_{11}), \nonumber \\
  H_3 &= -\frac{1}{4} ( C_{11-1} F_{11-1} Y_{1-1}
  + C_{110} F_{110} Y_{10} + C_{111} F_{111} Y_{11}),\nonumber \\
  H_4 &
  = \frac{1}{16} \Big(- \frac{A_{100} B_{100} D_{000}{}^2
    Y_{00}}{\pi^{1/2} W_{*}{}} + G_{11-1} Y_{1-1} + G_{110} Y_{10} +
  G_{111} Y_{11}\Big), \nonumber \\
  H_5 &=-\frac{1}{16} \Big( G_{11-1}
  Y_{1-1} + G_{110} Y_{10} + G_{111} Y_{11}\Big).
\end{align}

To obtain the physical bad field it is enough to recall
that~$\tilde{\phi}=\Theta \phi$ and rewrite equation
\eqref{Bad_unphysical} in terms of the physical coordinates. A
calculation then reveals the following:

\begin{proposition}
  \label{prop:bad}
  \emph{The solution~$\tilde{\phi}_b$ to equation~\eqref{gbu-good}
  ---the physical \emph{bad} equation--- arising from analytic initial
  data close to the cylinder at spatial infinity $I$ has the following
  formal expansion
  \begin{align}\label{Sol_Bad_From_Cylinder}
   \tilde{\phi}_b = & \frac{A_{100} B_{100} D_{000}{}^2 \log(\kappa )
     Y_{00}}{16 \pi^{1/2} W_{*}{} \tilde{v} \tilde{u}} +
   \frac{1}{\tilde{\rho}} \Big( C_{000} F_{000} Y_{00} + \tfrac{1}{2} G_ {000}
   \log(\kappa ) Y_{00} \nonumber \\ & + \frac{A_{100} (B_{100} D_{000}{}^2
     + 8
     \pi^{1/2} W_{*}{} F_{100}) Y_{00}}{16 \pi^{1/2} W_{*}{}
     \tilde{v}} + \frac{B_{100} (A_{100} D_{000}{}^2 -8 \pi^{1/2}
     W_{*}{} G_{100}) Y_{00}}{16 \pi^{1/2} W_{*}{} \tilde{u}} \Big)
   \nonumber\\
   & \frac{1}{\tilde{\rho}^2}
   \Big(C_{11-1} F_{11-1} Y_{1-1} + C_{110} F_{110} Y_{10} +
   C_{111} F_{111} Y_{11}\Big) \nonumber \\
   & + \frac{1}{32\tilde{\rho}^2}
   \Big(\frac{\tilde{u}}{\tilde{v}}+2\log\kappa
   -\frac{\tilde{v}}{\tilde{u}}
   \Big)
   (C_{11-1} F_{11-1} Y_{1-1} + C_{110} F_{110} Y_{10} + C_{111} F_{111} Y_{11})
   +\mathcal{O}(\tilde{\rho}^{-3})
  \end{align}
  }
\end{proposition}

%%%%%%%%%%%%%%%%%%%%%%%%%%%%%%%%%%%%%%%%%%%%%%%%%%%%%%%%%%%%%
\subsection{Comparison with the higher-order asymptotic expansion }
%%%%%%%%%%%%%%%%%%%%%%%%%%%%%%%%%%%%%%%%%%%%%%%%%%%%%%%%%%%%%

The most clean case to compare the logs obtained
in~\cite{DuaFenGasHil21} and those appearing through the analysis of
the cylinder at spatial infinity is that of the good field: on the one
hand, the asymptotic expansion for the good field~$\tilde{\phi}_g$
reported in~\cite{DuaFenGasHil21} does not contain log terms while the
solution of Proposition \ref{prop:good} contains logs. However, to go
beyond this obvious observation and to make a comparison in a more
equal footing it is necessary to recall that the expansions
in~\cite{DuaFenGasHil21} are obtained through integration along
outgoing null directions and the parameter along this curve was chosen
to be the physical radial coordinate~$\tilde{\rho}$. Therefore to put
the expansion~\eqref{Sol_Good_From_Cylinder} in the same format, one
needs to evaluate it at a fiduciary retarded time
\begin{align}\label{evaluation_u}
  \tilde{u}=-|\tilde{u}_\star|.
\end{align}
Hence on these surfaces one has
\begin{align}\label{param_with_radial}
  \tilde{v}= -|\tilde{u}_\star|+ 2 \tilde{\rho},
\end{align}
where~$\tilde{\rho}>|\tilde{u}_\star|/2$ so that~$\tilde{v}>0$
and~$\tilde{u}<0$.  Then, substituting
expressions~\eqref{evaluation_u} and~\eqref{param_with_radial} in the
expansion~\eqref{Sol_Good_From_Cylinder} one obtains, after Taylor
expanding close to the associated cut of null
infinity~$\mathcal{C}_\star \subset \mathscr{I}$, the following
expression:
\begin{flalign}\label{Sol_Good_From_Cylinder_Comparisson}
  \tilde{\phi}_g = & \frac{C_{000}Y_{00}}{\tilde{\rho}} +
  \frac{1}{16|\tilde{u}_\star|\tilde{\rho}}
  \Big( 8 B_{100}Y_{00} + D_{11-1}Y_{1-1} + D_{110}Y_{10} +D_{111}Y_{11 }\Big)
  + \frac{D_{000}Y_{00}}{2\tilde{\rho}}\log \bigg( \frac{2\tilde{\rho}}
  {|\tilde{u}_\star|}\bigg)
  \nonumber\\
  +& \frac{1}{32\tilde{\rho}^2} \Big( 8(A_{100}-|\tilde{u}_\star|D_{000})
  + (8C_{11-1}-D_{11-1})Y_{1-1} + (8C_{110}-D_{110})Y_{10}
  + (8C_{111}-D_{111})Y_{11 }    \Big)\nonumber \\
  + &\frac{1}{16\tilde{\rho}^2}\log
  \bigg( \frac{2\tilde{\rho}}{|\tilde{u}_\star|}\bigg)
  (D_{11-1}Y_{1-1} + D_{110}Y_{10} +D_{111}Y_{11 })
  + \mathcal{O}(\tilde{\rho}^{-3}\log{\tilde{\rho}}).
\end{flalign}

\begin{remark}\emph{
  The last expression contains terms proportional
  to~$\log(\tilde{\rho})$ for any finite~$\tilde{u}_{\star}\neq 0$.
  In contrast, the expression for the good field reported in
  \cite{DuaFenGasHil21} does not contain these logs.  Naturally,
  choosing initial data so that~$D_{p;\ell,m} = 0$ (the data
  for~$\dot{a}_{p;p,m}(0)$, see Remark~\ref{logfreeRemark}) one
  recovers a log-free expansion ---see
  Proposition~\ref{logfree_data_physgood}.}
  \end{remark}
  
Similarly, for the ugly field, using equations~\eqref{evaluation_u}
and~\eqref{param_with_radial} on the
expansion~\eqref{Sol_Ugly_From_Cylinder} gives:
\begin{align}\label{Sol_Ugly_From_Cylinder_Comparisson}
  & \tilde{\phi}_u =  \frac{2C_{000}Y_{00}}{\tilde{\rho}}  +
  \frac{-2B_{100}Y_{00} +  C_{11-1}  Y_{1-1} 
  + C_{110} Y_{10} +  C_{111}  Y_{11} }{4\tilde{\rho}^2} 
  - \frac{B_{100}Y_{00}}{2\tilde{\rho}^2}
  \log\bigg(\frac{|\tilde{u}_{\star}|}{2\tilde{\rho}}\bigg)
  + \mathcal{O}(\tilde{\rho}^{-3}\log \tilde{\rho}).
\end{align}
\begin{remark}
  \emph{To this order, the last expansion agrees with the general form
  expected for the expansion of the ugly field as given
  in~\cite{DuaFenGasHil21} since~$\log\tilde{\rho}$ starts from the
  second order in~$\tilde{\rho}$. Observe that the potential logs
  inherited from the good field ---and directly related to
  the~$i^0$-cylinder framework--- appearing
  equation~\eqref{Sol_Ugly_From_Cylinder_RAW} get removed after
  imposing initial data ensuring~$Q=0$ within the working
  Ansatz~\eqref{AnsatzGood} ---see Remark~\ref{Remark:ugly}. Hence,
  unlike the case of the good and bad fields where both contributions
  are present, the logs in~\eqref{Sol_Ugly_From_Cylinder_Comparisson}
  correspond exclusively to those reported in~\cite{DuaFenGasHil21}.
  }
  \end{remark}

Finally, proceeding analogously with the bad field using equation
\eqref{Sol_Bad_From_Cylinder}, one obtains:
\begin{flalign}\label{Sol_Bad_From_Cylinder_Comparisson}
  \tilde{\phi}_b = & \frac{1}{\tilde{\rho}}\Big[ C_{000} F_{000}
    Y_{00} -\frac{A_{100} B_{100} D_{000}{}^2Y_{00}}{16\pi^{1/2}W_{*}|
      \tilde{u}_{*}{}|} 
    \nonumber\\
    &\quad -\frac{1}{16 |\tilde{u}_{*}{}|}(
    8 B_{100} G_{100} Y_{00} + G_{11-1} Y_{1-1} + G_{110} Y_{10}
    + G_{111} Y_{11}) \Big] \nonumber \\ & +
  \frac{1}{64\tilde{\rho}^2}\Big[\frac{A_{100} (3 B_{100} D_{000}{}^2
      + 16 \pi^{1/2} W_{*}{} F_{100}) -16 \pi^{1/2} |\tilde{u}_{*}{}|
      W_{*}{} G_{000}) Y_{00}}{\pi^{1/2} W_{*}{}} \nonumber \\ & \qquad\quad
    + 2 ((8
    C_{11-1} F_{11-1} - G_{11-1}) Y_{1-1} + (8 C_{110} F_{110} -
    G_{110}) Y_{10} + (8 C_{111} F_{111} - G_{111}) Y_{11})\Big]
  \nonumber \\ &
  +\frac{1}{64\pi^{1/2}W_{*}}\log\Big(\frac{2\tilde{\rho}}
  {|\tilde{u}_\star|}\Big)
  \Big[ \frac{2(- A_{100} B_{100} D_{000}{}^2 + 16 \pi^{1/2}
      |\tilde{u}_{*}{}| W_{*}{} G_{000})
      Y_{00}}{|\tilde{u}_{*}{}|\tilde{\rho}} \nonumber \\ & \qquad \qquad
    - \frac{A_{100} B_{100} D_{000}{}^2 Y_{00}
      -4\pi^{1/2} W_{*}
      (G_{11-1} Y_{1-1} + G_{110} Y_{10} + G_{111} Y_{11})}
    {\tilde{\rho}^2}\Big]
  + \mathcal{O}(\tilde{\rho}^{-3}\log \tilde{\rho})
\end{flalign}

\begin{remark}
  \emph{The~$\log \tilde{\rho}$ terms in the last expression agree
  with the general form expected for the bad field according
  to~\cite{DuaFenGasHil21}. Nevertheless, the logarithmic terms in
  expression~\eqref{Sol_Bad_From_Cylinder_Comparisson} have two types
  of contributions: one contribution comes from the logs present in
  the expansion of the good field ---with initial data~$D_{p;\ell,m}
  \neq 0$--- and the other comes from terms which are present even for
  a log-free expansion of the good field ---with initial
  data~$D_{p;\ell,m} = 0$.}
  \end{remark}
  A reevaluation of the asymptotic system analysis to understand the
  apparent discrepancies is given in
  section~\ref{Asympt_sys_reevaluation}.

%%%%%%%%%%%%%%%%%%%%%%%%%%%%%%%%%%%%%%%%%%%%%%%%%%%%%%%%%%%%%
  \subsection{Revisiting the asymptotic system}
  \label{Asympt_sys_reevaluation}
%%%%%%%%%%%%%%%%%%%%%%%%%%%%%%%%%%%%%%%%%%%%%%%%%%%%%%%%%%%%%

The analysis of the previous section shows that the logarithmic terms
discussed in~\cite{DuaFenGasHil21} are not in correspondence with the
logarithmic terms having origin at spatial infinity using
the~$i^0$-cylinder framework. Since this discrepancy is more cleanly
shown by the expansion of the good field ---one method apparently
renders log-free expansions while the other does not--- it is of
interest to revisit the asymptotic approximation upon which the
expansions of~\cite{DuaFenGasHil21} were derived ---namely
H\"ormander's asymptotic system--- under the light of
Proposition~\ref{prop:good}.

H\"ormander's asymptotic system is based on the observation that
derivatives tangent to the outgoing null cone decay faster than
transverse derivatives to it. This leads to calling~$\underline{L}$
the bad derivative and, along with angular derivatives, calling~$L$
the good derivative. The first order asymptotic system for the wave
equation is based on the heuristic approach of disregarding the terms
that only contain good derivatives.  Expressing the physical wave
operator in Minkowski spacetime as in equation~\eqref{PhysWaveOp},
discarding the second term as it only contains two good ---angular---
derivatives, one obtains
\begin{align}
  L\underline{L}(\tilde{\rho}\tilde{\phi}) \simeq 0
\end{align}
where the symbol~$\simeq$ is used to emphasise that the previously
described asymptotic approximation has been used. The last expression
can be integrated as follows
\begin{align}\label{asymptsysSol}
  \partial_{\tilde{u}}\partial_{\tilde{v}}(\tilde{\rho}\tilde{\phi})
  \simeq & \;0,
  \nonumber\\ \partial_{\tilde{v}}(\tilde{\rho}\tilde{\phi}) \simeq &
  (\partial_{\tilde{v}}(\tilde{\rho}\tilde{\phi}))|_{\tilde{u}_\star}
  :=f(\tilde{v},\theta^A),
  \nonumber \\ \tilde{\rho}\tilde{\phi} \simeq &
  (\tilde{\rho}\tilde{\phi})|_{\tilde{v}_\star} +
  \int_{\tilde{v}_\star}^{\tilde{v}}f(\bar{v},\theta^A)d\bar{v},
  \nonumber \\ \implies \tilde{\phi} \simeq &
  \frac{1}{\tilde{\rho}}(G(\tilde{u},\theta^A)+F(\tilde{v},\theta^A)),
\end{align}
where in this calculation~$f$ appears as an ``integration constant''
determined from given data
for~$\partial_{\tilde{v}}(\tilde{\rho}\tilde{\phi})$
and~$G(\tilde{u},\theta^A)$ is an ``integration constant'' resulting
from data for~$\tilde{\rho}\tilde{\phi}$.  Similarly the
function~$F(\tilde{v},\theta^A)$ is a shorthand
for~$\int_{\tilde{v}}f$ appearing in the third line of the derivation
of expression~\eqref{asymptsysSol}. Therefore, this approach can serve
as the foundation of a heuristic method for determining the general
form of the field in general asymptotically flat backgrounds as was
done in~\cite{DuaFenGasHil21}. Nonetheless, observe that the
functional form of~$F(\tilde{v},\theta^A)$ and~$G(\tilde{u},\theta^A)$
is not given explicitly by the method and are determined implicitly by
initial data, given for instance on~$\mathcal{S}$ when one considers
the Cauchy problem. Hence,~$F(\tilde{v},\theta^A)$
and~$G(\tilde{u},\theta^A)$ can accommodate for the logarithmic terms
appearing in Proposition~\ref{prop:good}. In other words, the
logarithmic terms arising from the critical sets are, in some sense,
not missed by the asymptotic system heuristics as they are contained
inside the~``integration constants''.  However, the asymptotic system
method by itself does not give information on the functional form of
these~``integration constants'' or the field itself at the critical
sets where spatial and null infinity meet. Nonetheless, one can still
retrieve the \emph{first order logarithmic term} appearing in the good
field by means of the leading order asymptotic system heuristics. To
do so notice that if initial data is chosen so that
\begin{align}
  f(\tilde{v},\theta^A)\simeq \tilde{v}^{-1}M(\theta^A) + o(\tilde{v}^{-1}),
\end{align}
where~$M$ is a function of~$\theta^A$ and the standard little-o
notation is used in the second term, then integrating the second line
in equation~\eqref{asymptsysSol} gives
\begin{align}
  \tilde{\phi} \simeq \frac{\log \tilde{v}}{\tilde{\rho}}M(\theta^A)
  + \frac{1}{\tilde{\rho}}G(\tilde{u},\theta^A),
\end{align}
Then, on outgoing null directions~$u=u_{\star}$ one recovers
\begin{align}
\tilde{\phi} \simeq \frac{\log
  \tilde{\rho}}{\tilde{\rho}}M( \theta^A) + \frac{1}{\tilde{\rho}}
G(\tilde{u},\theta^A),
\end{align}
Notice that these logarithmic terms were discarded in the analysis
of~\cite{DuaFenGasHil21}. Additionally, observe that the condition
ensuring the absence of the leading logarithmic term is
\begin{align}\label{ASnologleading}
  f(\tilde{v},\theta^A)\simeq \mathcal{O}(\tilde{v}^{-2}).
\end{align}
A attractive feature of expressing the ``no-leading-order-log
condition'' in terms of~\eqref{ASnologleading} is that it allows for a
simple \emph{physical interpretation}: if the \emph{initial data for
the incoming characteristic
variable}~$\partial_{\tilde{v}}(\tilde{\rho}\tilde{\phi})$ decays
faster than~$\mathcal{O}(\tilde{v}^{-1})$, then there will be no
leading log-term in the solution for~$\tilde{\phi}$
towards~$\mathscr{I}^{+}$. On the other hand, the analysis using
the~$i^0$-cylinder predicts a \emph{hierarchy of logarithmic terms} in
the expansion for the good field. The first order logarithmic term
obtained through the conformal approach is controlled at the level of
initial data by the coefficient~$D_{000}$ ---higher order logs being
controlled by~$D_{n,n,m}$ with~$n\geq 1$. Thus, verifying that the
asymptotic system heuristics described above capture the leading
log-term predicted by the~$i^0$-cylinder method one needs to check
their correspondence at the level of initial data. To do so, observe
that
\begin{align}
  f(\tilde{v},\theta):=
  (\partial_{\tilde{v}}(\tilde{\rho}\tilde{\phi}))|_{\tilde{u}_\star}=
  \big(\partial_{\tilde{t}}(\tilde{\rho}\tilde{\phi}) +
  \partial_{\tilde{\rho}}(\tilde{\rho}\tilde{\phi})\big)|_{\mathcal{S}},
\end{align}
thus, assuming that the initial data has a series expansion in integer
powers of~$\tilde{\rho}^{-1}$, a necessary and sufficient condition to
get rid of the leading log at null infinity from the point of view of
the asymptotic system heuristics is that
\begin{align}\label{nologleadID}
\big(\partial_{\tilde{t}}(\tilde{\rho}\tilde{\phi}) +
\partial_{\tilde{\rho}}(\tilde{\rho}\tilde{\phi})\big)|_{\mathcal{S}}
\sim \mathcal{O}(\tilde{\rho}^{-2}),
\end{align}
where to re-express \eqref{ASnologleading} it was used that
on~$u=u_{\star}$ one has $\tilde{v} \sim \tilde{\rho}$.  Rewriting the
condition~\eqref{nologleadID} in terms of the unphysical fields, using
the results from Proposition~\ref{Prop:FframeToPhsyicalframe} ---as
similarly done in Remark~\eqref{remarkID}--- and substituting the
initial data Ansatz~\eqref{ID_good}, one gets
\begin{align}\label{nologleadIDPhysUnphysRelation}
\big(\partial_{\tilde{t}}(\tilde{\rho}\tilde{\phi}) +
\partial_{\tilde{\rho}}(\tilde{\rho}\tilde{\phi})\big)|_{\mathcal{S}}
\simeq \frac{1}{\tilde{\rho}}D_{000} + \mathcal{O}(\tilde{\rho}^{-2}).
\end{align}
Hence the no-leading-log condition~$D_{000}=0$ obtained from
the~$i^0$-cylinder analysis corresponds precisely to the
condition~\eqref{nologleadID} expressed in the physical picture.
Therefore, one can conclude that the leading log obtained through the
above described method based on the asymptotic system retrieves the
first order log obtained through the~$i^0$-cylinder analysis.  Notice
however that the leading log corresponds to the spherically symmetric
solution (terms with~$Y_{00}$) while the higher order logs are related
to a specific harmonics in the Ansatz~\eqref{AnsatzGood} ---associated
to non-trivial spherical harmonics~$Y_{\ell m}$. Obtaining a
generalisation of the present new take on the asymptotic system
heuristics to higher order is left for future work.

Choosing initial data of compact support all log-terms are of course
suppressed. The foregoing discussion shows that the first order
asymptotic system captures the case in which log-terms are absent in
the initial data but manifest at future null infinity. But one
furthermore observes from the general solution to the asymptotic
system~\eqref{asymptsysSol} that examples in which log-terms {\it
  appear} in the initial data, but not at future null infinity are
easily constructed (mutatis mutandis at past null infinity). For
instance, one may choose,
\begin{align}
  \tilde{\phi}&=
  \frac{M_F(\theta^A)}{\tilde{\rho}}
  +\frac{H_G(\theta^A)\log|\tilde{u}|}{\tilde{\rho}},
\end{align}
where it is stressed that~$\tilde{u}$ is finite at any cut of future
null infinity. A final interesting case is that in which
logarithmically divergent terms are present both in the initial data
and at future null infinity, for instance
\begin{align}
  \tilde{\phi}&=
  \frac{H_F(\theta^A)\log|\tilde{v}|}{\tilde{\rho}}+
  \frac{M_G(\theta^A)}{\tilde{\rho}}.
\end{align}
The proposed discriminating condition for the appearance of leading
log-terms at future null infinity~\eqref{nologleadID} is compatible
with all four cases.

%%%%%%%%%%%%%%%%%%%%%%%%%%%%%%%%%%%%%%%%%%%%%%%%%%%%%%%%%%%%%
\section{Conclusions}
%%%%%%%%%%%%%%%%%%%%%%%%%%%%%%%%%%%%%%%%%%%%%%%%%%%%%%%%%%%%%

The peeling property of the gravitational field has been a continuous
source of debate in the general relativity community and a good number
of works on the topic have been presented in recent
years~\cite{DuaFenGas22, Lin17, Keh21, GasVal18, Fri18}. Nonetheless,
it is usually the case that these results are obtained in different
gauges, making different assumptions on the initial data and using
different formulations of the Einstein field equations. This paper is
a first step into understanding the relation, or the lack thereof,
between the polyhomogeneous expansions obtained in~\cite{DuaFenGas22}
and those in~\cite{Fri98a, Val07, GasVal18}. Although both expansions
give rise to polyhomogeneous terms in the Weyl scalars, they are
obtained with strikingly different formulations of the Einstein field
equations and gauges. On the one hand, the formulation
in~\cite{DuaFenGas22, DuaFenGasHil22a} is based on a hyperbolic
reduction of the Einstein field equations in generalised harmonic
gauge so that the central variables are the components of the
(physical) spacetime metric. On the other hand the formulation under
which~\cite{Fri98a, Val07, GasVal18} are derived is a curvature
oriented formulation of the conformally extended (unphysical)
spacetime where the gauge (called the~$F$-gauge) is fixed through a
congruence of conformal geodesics.

Since the basic strategy to obtain the polyhomogeneous expansions for
the Weyl scalars in~\cite{DuaFenGas22} is based on the method
of~\cite{DuaFenGasHil21} which was developed taking as a base case the
analysis of a model of equations known as the GBU model, then it seems
natural to investigate the same model system using the methods of the
cylinder at spatial infinity. Therefore, analysing the GBU model in
Minkowski spacetime where the relation between the two gauges can be
written in closed form represents an opportunity to make a clear-cut
comparison of the logarithmic terms appearing by using each method.
The conclusion of this comparison here is that the logarithmic terms
presented in~\cite{DuaFenGasHil21} and those using the framework of
the cylinder at spatial infinity are not the same. The clearest case
for comparison is the good field, where the~$i^0$-cylinder approach
shows a polyhomogeneous expansion close to~$i^0$. These logarithmic
terms can be avoided if special initial data with $D_{p;\ell,m}=0$ is
chosen.  For the case of the ugly and bad fields the logarithmic terms
appear at the order expected from the analysis
of~\cite{DuaFenGasHil21}.  For the bad field an analogous observation
can be made, there are, however, two contributions to the logarithmic
terms as it can be seen from the coefficients in the expansion: one
contribution is inherited from the logarithmic terms of the expansion
of the good field ---associated with initial data
with~$D_{p;\ell,m}\neq 0$--- and the other comes from an integration
used to construct the solution. Hence, the question is what was missed
by the asymptotic system analysis employed and extended to higher
orders in~\cite{DuaFenGasHil21}? In order to answer this question the
original first order asymptotic system for the good equation ---the
one that more clearly shows the difference--- was revisited and it was
discussed how the missing logs are contained inside the ``integration
constants'' generated by the method. These integration constants are
in fact functions of either~$(\tilde{u},\theta^A)$
or~$(\tilde{v},\theta^A)$, and inherited on null hypersurfaces from
Cauchy data. Hence, the asymptotic system method itself does not give
information about the form of these functions close to spatial
infinity. Exploiting that these functions are determined by the
initial data it was shown that the first-order logarithmic term for
the good field can be retrieved using the asymptotic system
heuristics, and it was shown this term indeed corresponds to the
leading log-term obtained using the conformal~$i^0$-cylinder
method. Moreover, this calculation allowed to give a physical
interpretation to the first-order no-log condition in terms of the
decay of the data incoming characteristic
variable~$\partial_{\tilde{v}}(\tilde{\rho}\tilde{\phi})$. Furthermore,
it was shown, within the asymptotic system, that there is no logical
implication between the presence of leading order log-terms in initial
data and at null infinity. Whether this discussion can be extended to
recover the \emph{full hierarchy} (higher-order) of logarithmic terms
obtained using~$i^0$-cylinder method is left for future work.

In the discussion given in~\cite{DuaFenGas22} it was shown that the
violation of peeling by the logarithmic terms arising from the method
laid out in~\cite{DuaFenGasHil21} can be avoided, hence retrieving the
classical peeling result, by suitably choosing gauge source functions
and adding multiples of the constraints to the evolution equations. In
other words, the logarithmic terms in~\cite{DuaFenGasHil21} are
gauge. It should be stressed that expansions obtained through the
asymptotic systems approach and conformal methods are, at the time of
writing, still formal in the sense that rigorous PDE estimates have
not been developed for the full non-linear equations so far. However,
it is the general expectation that the logarithmic terms originating
at the critical sets~$I^{\pm}$ given in~\cite{Fri98a, FriKan00, Fri18,
  GasVal18} are {\it not} gauge and hence cannot be removed. Whether
the latter expectation is justified is yet to be confirmed.

%%%%%%%%%%%%%%%%%%%%%%%%%%%%%%%%%%%%%%%%%%%%%%%%%%%%%%%%%%%%%
\subsection*{Acknowledgements}
%%%%%%%%%%%%%%%%%%%%%%%%%%%%%%%%%%%%%%%%%%%%%%%%%%%%%%%%%%%%%
 
We have profited from scientific discussions and interaction in the
online \emph{Conformal/spinorial workshop} lead by Juan Valiente
Kroon. EG holds a FCT (Portugal) investigator grant
2020.03845.CEECIND. DH acknowledges support from the PTDC/MAT-
APL/30043/2017. JF acknowledges support from FCT (Portugal) programs
PTDC/MAT-APL/30043/ 2017, UIDB/00099/ 2020. MD acknowledges support
from FCT (Portugal) program PD/BD/135511/2018.

%%%%%%%%%%%%%%%%%%%%%%%%%%%%%%%%%%%%%%%%%%%%%%%%%%%%%%%%%%%%%
\normalem
\bibliographystyle{unsrt}
\bibliography{GBU_cylinder_i0}
%%%%%%%%%%%%%%%%%%%%%%%%%%%%%%%%%%%%%%%%%%%%%%%%%%%%%%%%%%%%%

%%%%%%%%%%%%%%%%%%%%%%%%%%%%%%%%%%%%%%%%%%%%%%%%%%%%%%%%%%%%%
\end{document}